\documentclass[fleqn,usenatbib,useAMS]{mnras}
\usepackage{amssymb,amsmath}
\usepackage{graphicx}
\usepackage{xcolor,natbib}
\usepackage{multirow}
\usepackage[T1]{fontenc}
\usepackage{ae,aecompl}
\usepackage{newtxtext, newtxmath}
\usepackage{float}
\usepackage{subfigure}
\usepackage{longtable}
\usepackage{enumitem}
\usepackage{longtable,listings}
\usepackage[flushleft]{threeparttable}
\usepackage{parcolumns}
\def\dif{\mathop{}\!\mathrm{d}}
\def\obj{SDSS J1257+2023}

\title[Different Broad Balmer Lines]{A candidate of binary black hole system in AGN with broad Balmer emission lines 
having quite different line widths}
\author[Zhang X. G.]{XueGuang Zhang$^{1}$
\thanks{Contact e-mail: \href{mailto:aexueguang@qq.com}{aexueguang@qq.com}}\\
$^{1}$School of Physical Science and Technology, Guangxi University, No. 100, Daxue East Road, Nanning, 530004, 
P. R. China}

\begin{document}
\label{firstpage}
\pagerange{\pageref{firstpage}--\pageref{lastpage}}

\maketitle

\begin{abstract} %%%about 249 words
	In the manuscript, a candidate of sub-pc binary black hole (BBH) system is reported in SDSS J1257+2023 through 
different properties of broad Balmer emission lines. After subtractions of host galaxy contributions, Gaussian 
functions are applied to measure emission lines in SDSS J1257+2023, leading line width (second moment) 760${\rm km/s}$ 
of broad H$\beta$ to be 0.69 times of line width 1100${\rm km/s}$ of broad H$\alpha$, quite different from normal 
line width ratio 1.1 of broad H$\beta$ to broad H$\alpha$ in quasars. The quite broader component in broad H$\alpha$ 
in SDSS J1257+2023 can be confirmed with confidence level higher than $5\sigma$ through F-test technique, through 
different model functions applied to measure emission lines. The broad Balmer emission lines having different line 
widths can be naturally explained by a BBH system with different obscurations on central two independent BLRs. 
Meanwhile, through ZTF light curves and corresponding phase folded light curves well described by sinusoidal function, 
BBH system expected optical QPOs can be detected with periodicity about 1000days, confirmed with confidence level 
higher than $3\sigma$ by Generalized Lomb-Scargle periodogram. And through CAR process simulated light curves, 
confidence level higher than $2\sigma$ can be determined to support the optical QPOs in SDSS J1257+2023 not from 
intrinsic AGN activities, although the ZTF light curves have short time durations. Moreover, through oversimplified 
BBH system simulated results, studying different broad Balmer lines as signs of BBH systems in normal quasars with 
flux ratios around 4 of broad H$\alpha$ to broad H$\beta$ could be done in near future.
\end{abstract}

\begin{keywords}
galaxies:active - galaxies:nuclei - quasars:emission lines - quasars:individual (SDSS J1257+2023)
\end{keywords}

\section{Introduction}

%%first

	Dual galactic core systems (dual galaxy systems, or dual AGN systems, or galaxy-AGN pair systems in literature) 
with space separations of hundreds to thousands of parsecs (pcs) to supermassive binary black hole (BBH) systems with space 
separations of sub-parsecs are commonly expected in galaxies as well discussed in \citet{bb80, mk10, fg19, mj22}, after 
considering galaxy merging as one of fundamental and essential processes of galaxy formation and evolution as discussed 
in \citet{cr92, lc93, kw93, bh96, sr98, mh01, lk04, md06, bf09, se14, rp16, rs17, bh19, mj21, yp22}. And different 
techniques have been proposed to detect dual core systems and/or BBH systems, such as applications of double-peaked 
features of broad and/or narrow emission lines as signs of dual core systems and/or BBH systems in \citet{zw04, kz08, 
bl09, ss09, sl10, eb12, pl12, cs13, le16, wg17}, through spatially resolved central regions of galaxies in \citet{km03, 
rt09, pv10, ne17, kw20}, applications of long-standing optical Quasi-Periodic Oscillations (QPOs) with periodicities of 
hundreds to thousands of days as signs of BBH systems in \citet{gd15a, gd15b, zb16, lg18, kp19, ss20, lw21, zh22a}, etc.. 
More recently, \citet{zh21d} have reported the different broad Balmer emission line features as the sign to support a 
central BBH systems in SDSS J1547 with double-peaked broad H$\beta$ but single-peaked broad H$\alpha$. More 
recent reviews on dual core systems and BBH systems can be found in \citet{dv19, ch22}.

	Commonly, one dual core system at kilo-parsec scales should lead to no effects of central spatial structure of 
one core on broad emission line properties related to the other core, also lead to rotating time scale longer than $10^6$ 
years. Meanwhile, one BBH system at sub-parsec scales can well not only lead to apparent effects of central spatial 
structure of one core on broad emission line properties related to the other core, but also lead to rotating time scale 
around hundreds to thousands of days. In the manuscript, not only QPOs in long-term variabilities but also special 
properties of broad emission lines should be applied, therefore, rather than dual core systems at kilo-parsec scales, 
BBH systems at sub-parsecs scales are mainly considered in the manuscript.

	Among the proposed techniques to detect BBH systems, applications of different properties of broad Balmer 
emission lines are very interesting. Broad Balmer lines from central broad emission line regions (BLRs) are the 
apparently strongest broad lines in optical band of broad line AGN (Type-1 AGN) \citep{ks00, sm00, kg04, pe04, hc08, 
bw10, sh11, bl21}. Considering totally similar emission regions, optical broad Balmer lines have similar line profiles. 
Even considering different optical depths of Balmer emission lines, the Balmer lines have also similar line profiles, 
such as the shown results in \citet{gh05} that line widths of broad H$\beta$ are strongly linearly correlated with 
those of broad H$\alpha$ in a large sample of SDSS (Sloan Digital Sky Survey) quasars. Moreover, as the shown composite 
spectrum of SDSS quasars in \citet{vd01}, there are also similar line profiles of broad Balmer emission lines. More 
recently, \citet{nr22} have studied kinematic properties of emission regions of broad H$\alpha$ and broad H$\beta$ 
through a large sample of Type-1 AGN with high quality spectra, to confirm that the broad H$\beta$ and broad H$\alpha$ 
line emission gases follow similar kinematics, to support similar line profiles of broad Balmer emission lines in 
Type-1 AGN. However, considering BBH systems, there are two independent BLRs with probably different kinematic or 
physical properties related to central two active nuclei, which will easily lead to different properties 
of broad Balmer emission lines due to different contributions from different BLRs, to support different broad Balmer 
lines as signs of BBH systems, as reported and discussed in \citet{zh21d}.

	Among the SDSS quasars, it is hard to find the second target which has apparent double-peaked broad H$\beta$ 
but apparent single-peaked broad H$\alpha$. However, among the SDSS quasars, an interesting target, the quasar SDSS 
J125741.17+202347.80 (=\obj) at redshift 0.081, is found and reported in the manuscript as a new BBH system candidate, 
due to its quite different properties of single-peaked broad Balmer emission lines and further clues from probable 
optical QPOs through long-term variabilities from Zwicky Transient Facility (ZTF) \citep{bk19, ds20}. Section 2 presents 
the main spectroscopic results on different properties of broad H$\alpha$ and broad H$\beta$. Section 3 gives long-term 
variability properties and necessary discussions. Section 4 shows the further applications. Section 5 gives final 
summary and conclusions. In the manuscript, the cosmological parameters have been adopted as 
$H_{0}=70{\rm km\cdot s}^{-1}{\rm Mpc}^{-1}$, $\Omega_{\Lambda}=0.7$ and $\Omega_{\rm m}=0.3$.

\begin{figure*}
\centering\includegraphics[width = 8cm,height=5cm]{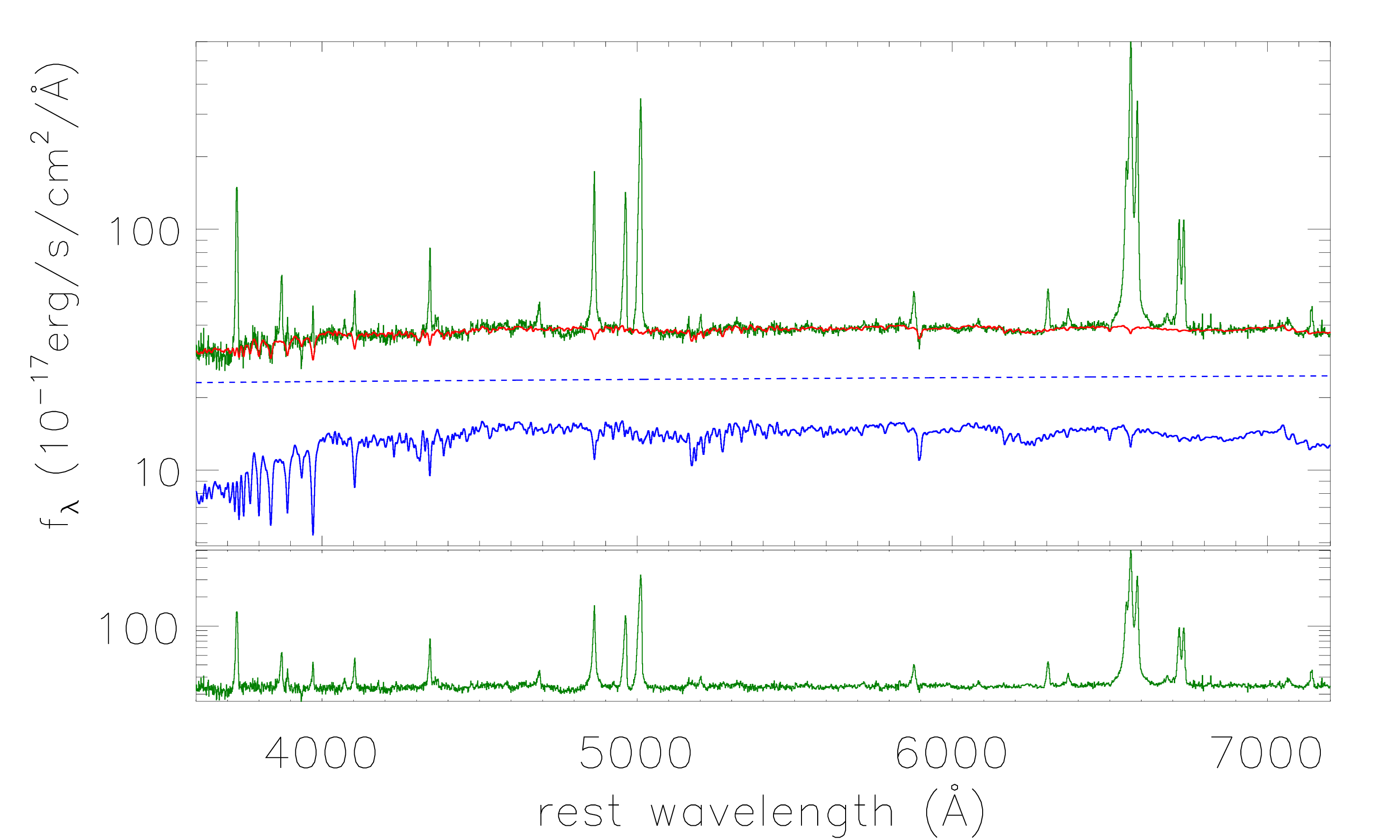}
\centering\includegraphics[width = 8cm,height=5cm]{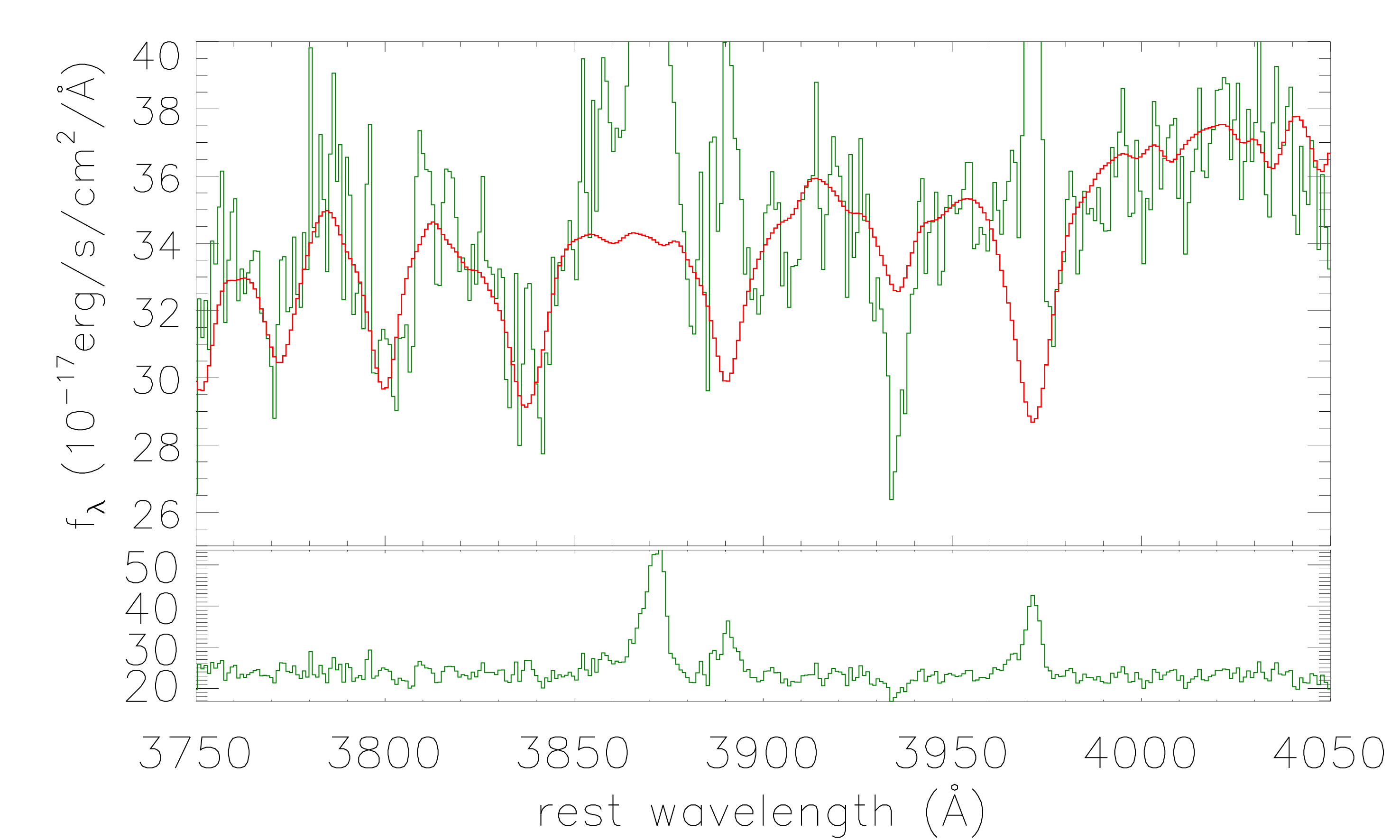}
\caption{Top left panel shows the SSP method determined descriptions (solid red line) to the SDSS spectrum (solid dark green 
line) with emission lines being masked out. In top left panel, solid blue line and dashed blue line show the determined host 
galaxy contributions and power law AGN continuum emissions, respectively. Bottom left panel shows the line spectrum calculated 
by the SDSS spectrum minus the host galaxy contributions. Here, due to strong emission lines in \obj, the Y-axis is in 
logarithmic coordinate, in order to show clear host galaxy absorption features in top left panel and to show clear emission 
lines in the line spectrum in bottom left panel. Right panels show corresponding results for the SDSS spectrum 
within rest wavelength from 3750\AA~ to 4050\AA, in order to clearly show best fitting results to absorption features around 
4000\AA.}
\label{spec}
\end{figure*}

\begin{figure*}
\centering\includegraphics[width = 8cm,height=5cm]{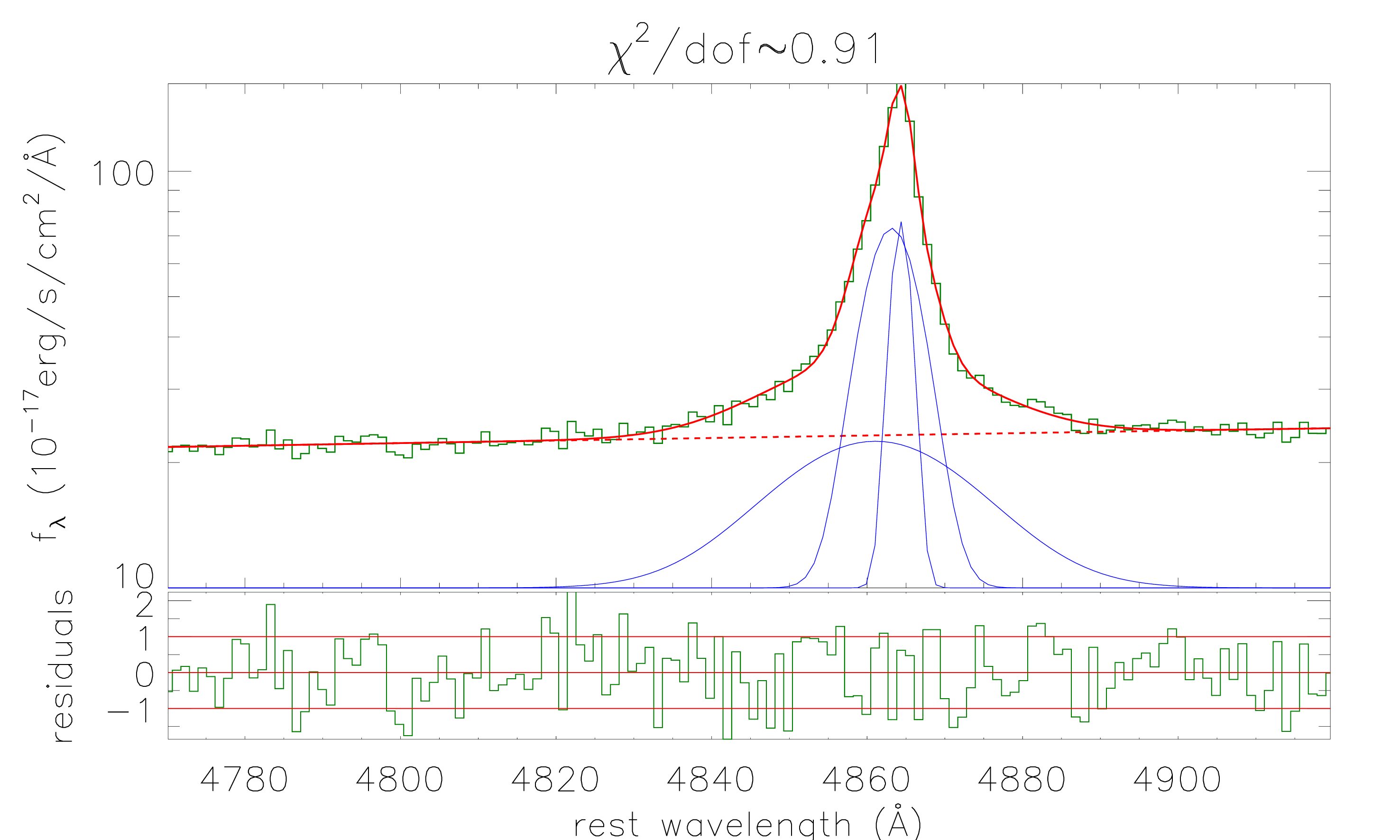}
\centering\includegraphics[width = 8cm,height=5cm]{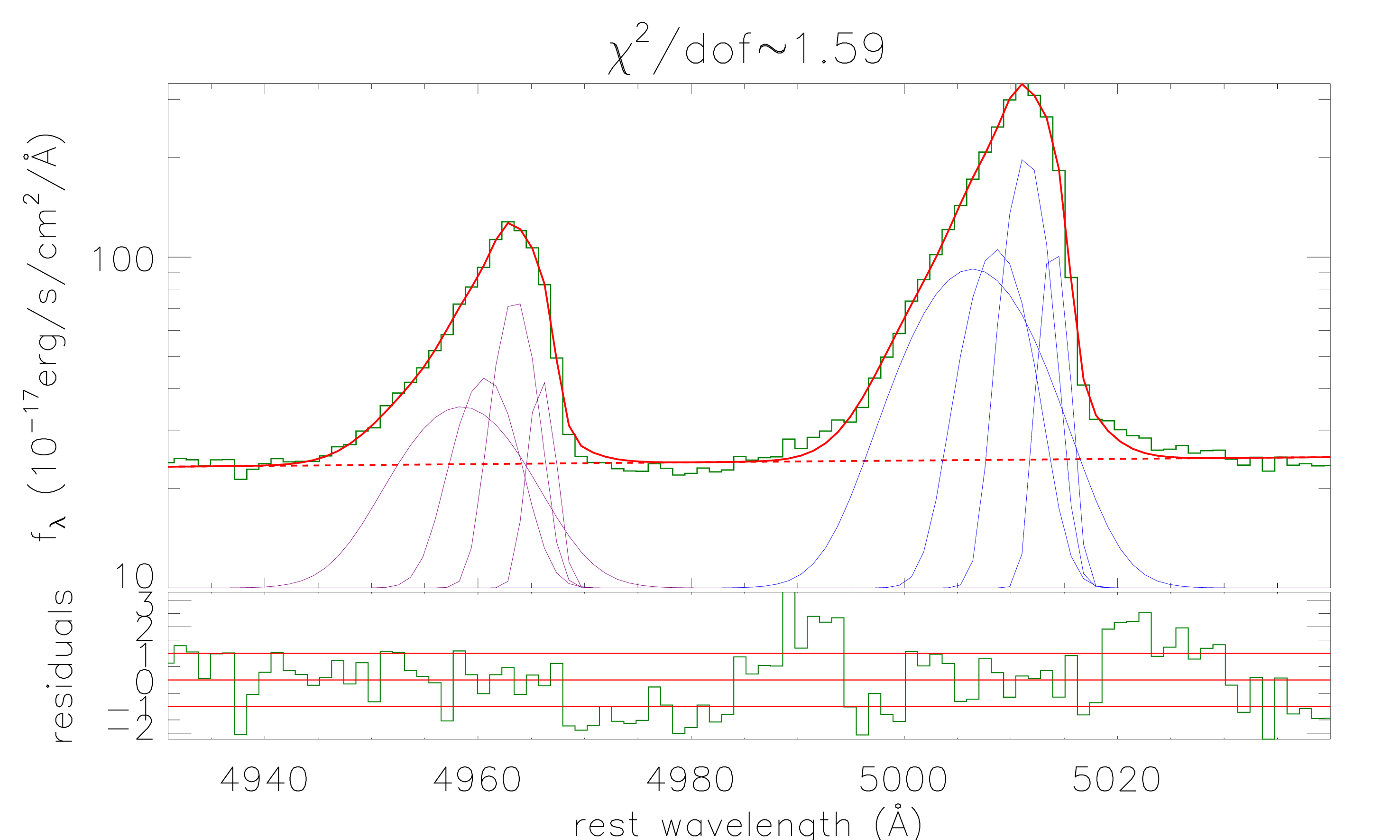}
\centering\includegraphics[width = 8cm,height=5cm]{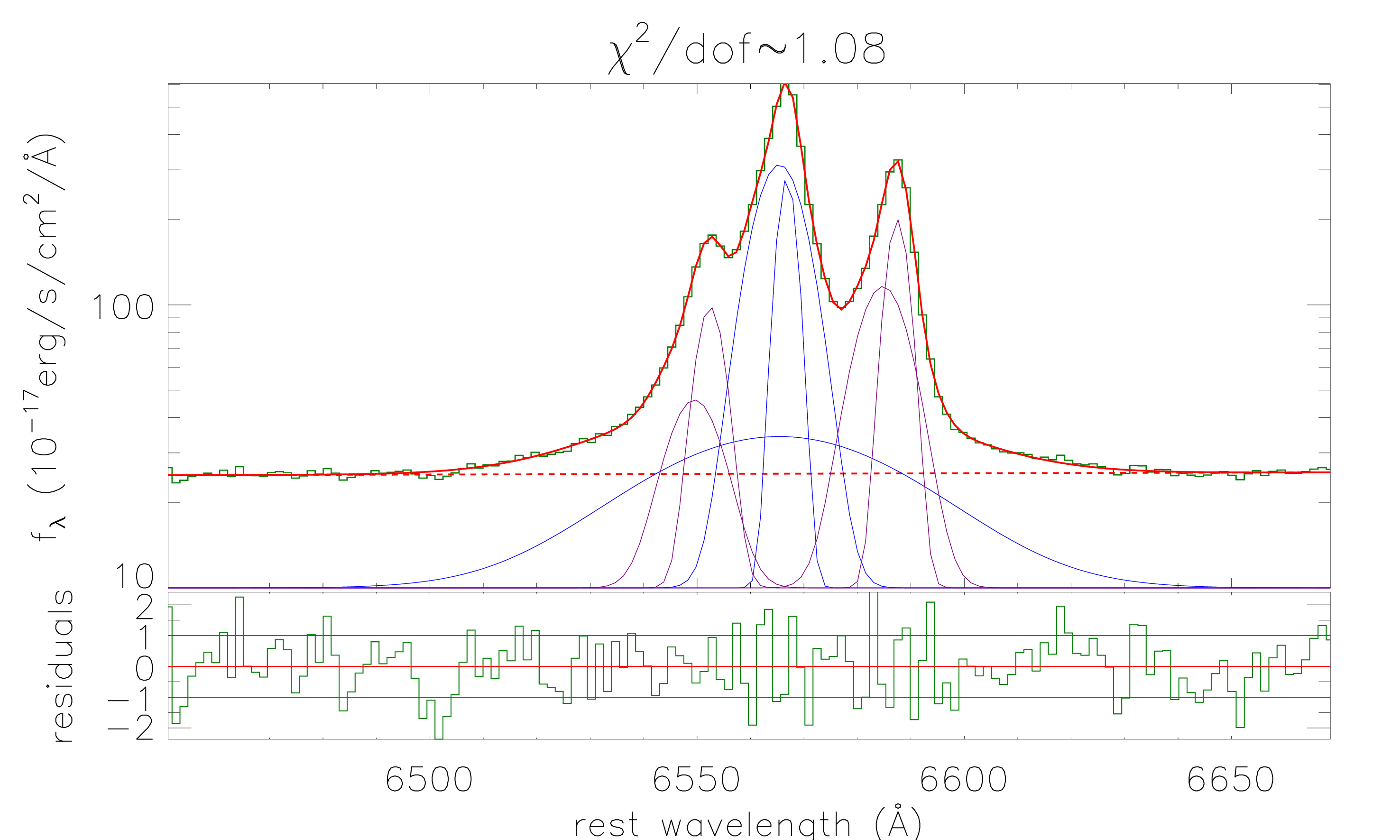}
\centering\includegraphics[width = 8cm,height=5cm]{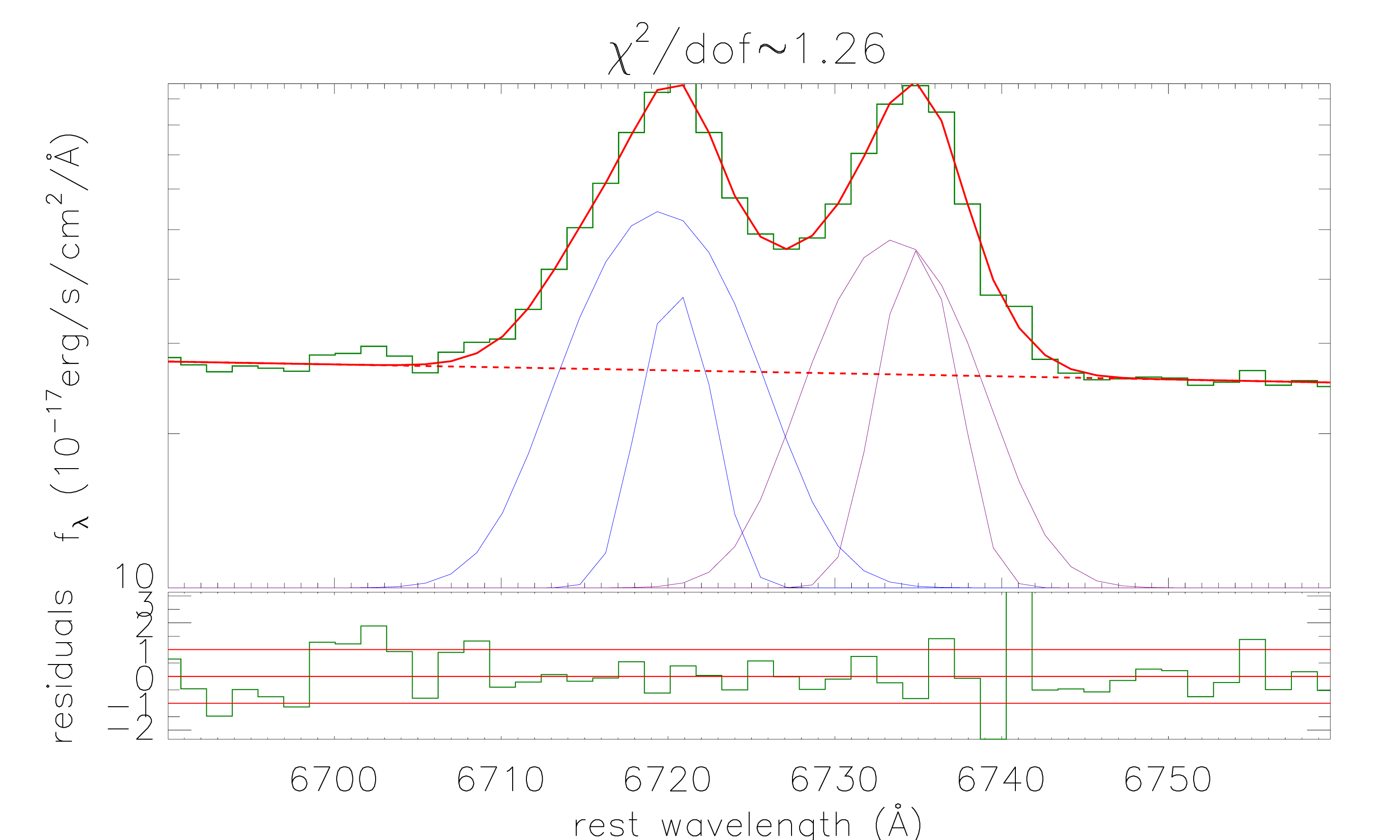}
\caption{Top left panel shows the best-fitting results (top region) and corresponding residuals (bottom region) to the 
H$\beta$ emission line. Top right panel, bottom left panel and bottom right panel show corresponding results for the 
[O~{\sc iii}] doublet, for the H$\alpha$ and [N~{\sc ii}] doublet, and for the [S~{\sc ii}] doublet, respectively. In 
top region of each panel, solid dark green line shows the line spectrum after subtractions of host galaxy contributions, 
solid red line shows the best fitting results. In bottom region of each panel for residuals, horizontal red lines show 
residuals=$0,~\pm1$, respectively. In top region of top left panel, solid blue lines show the determined three Gaussian 
components in H$\beta$, one for broad H$\beta$ and two for narrow H$\beta$. In top region of top right panel, solid blue 
lines and solid purple lines show the determined Gaussian components in [O~{\sc iii}] doublet. In top region of bottom 
left panel, solid blue lines show the determined three Gaussian components (one for broad H$\alpha$ and two for narrow 
H$\alpha$) in H$\alpha$, and solid purple lines show the determined Gaussian components in [N~{\sc ii}] doublet. In top 
region of bottom right panel, solid blue lines and solid purple lines show the determined Gaussian components in 
[S~{\sc ii}] doublet. The determined $\chi^2/dof$ is marked in title of each panel. }
\label{ha}
\end{figure*}

\iffalse
\begin{figure}
\centering\includegraphics[width = 8cm,height=5cm]{o31.ps}
\centering\includegraphics[width = 8cm,height=5cm]{o32.ps}
\caption{Fitting results to [O~{\sc iii}] doublet by three Gaussian functions applied to each line (top panel) and by 
two Gaussian functions applied to each line (bottom panel). Symbols and line styles have the same meanings as those 
in top right panel of Fig.~\ref{ha}.
}
\label{o32}
\end{figure}
\fi

\begin{figure}
\centering\includegraphics[width = 8cm,height=5cm]{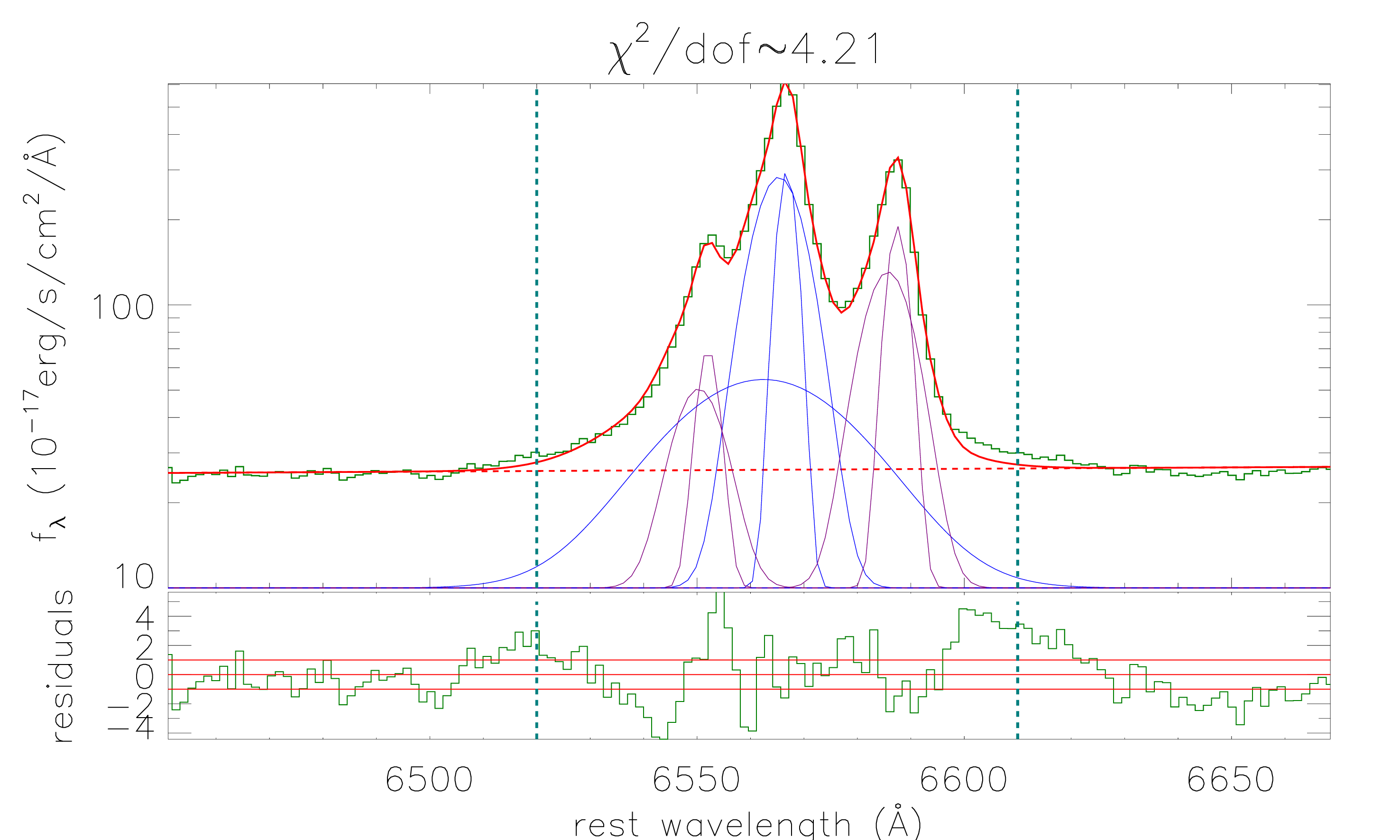}
\centering\includegraphics[width = 8cm,height=5cm]{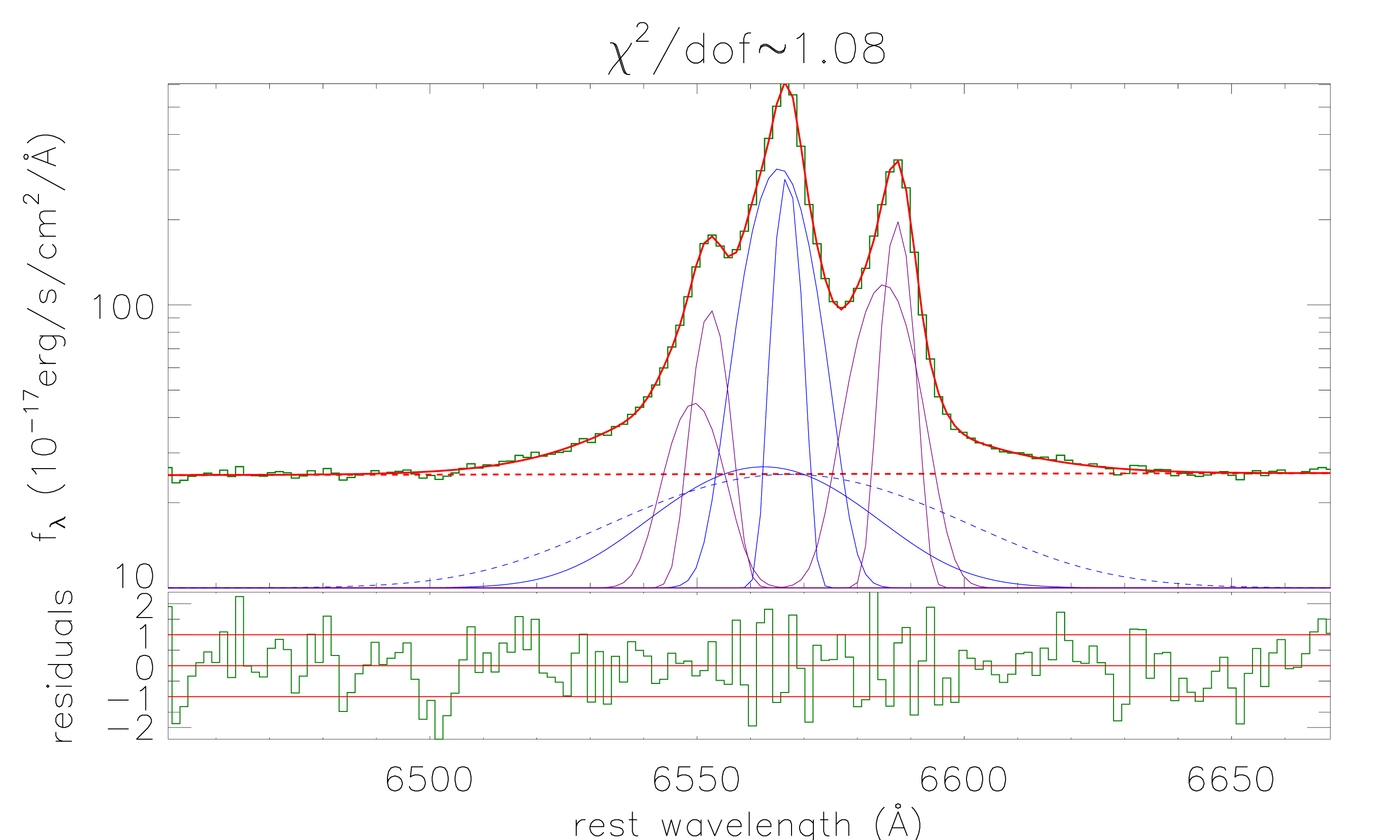}
\caption{Top panel shows the determine descriptions (top region) and corresponding residuals (bottom region) to emission 
lines around H$\alpha$ after considering H$\alpha$ having the same line profile (the same redshift and the same second 
moment in the three Gaussian components) of H$\beta$. Bottom panel shows the best descriptions (top region) and 
corresponding residuals (bottom region) to emission lines around H$\alpha$ after considering H$\alpha$ described by 
similar profile of H$\beta$ plus an additional broad Gaussian component. In top panel, vertical dashed cyan lines mark 
positions for bad descriptions to the wings around the emission lines, considering the same line profiles of H$\alpha$ 
and H$\beta$, the other line styles have the same meanings as those shown in bottom left panel of Fig.~\ref{ha}. In bottom 
panel, dashed blue line shows the determined additional broad Gaussian component in H$\alpha$, the other line styles have 
the same meanings as those shown in bottom left panel of Fig.~\ref{ha}.}
\label{ha2}
\end{figure}

\begin{figure*}
\centering\includegraphics[width = 16cm,height=7.75cm]{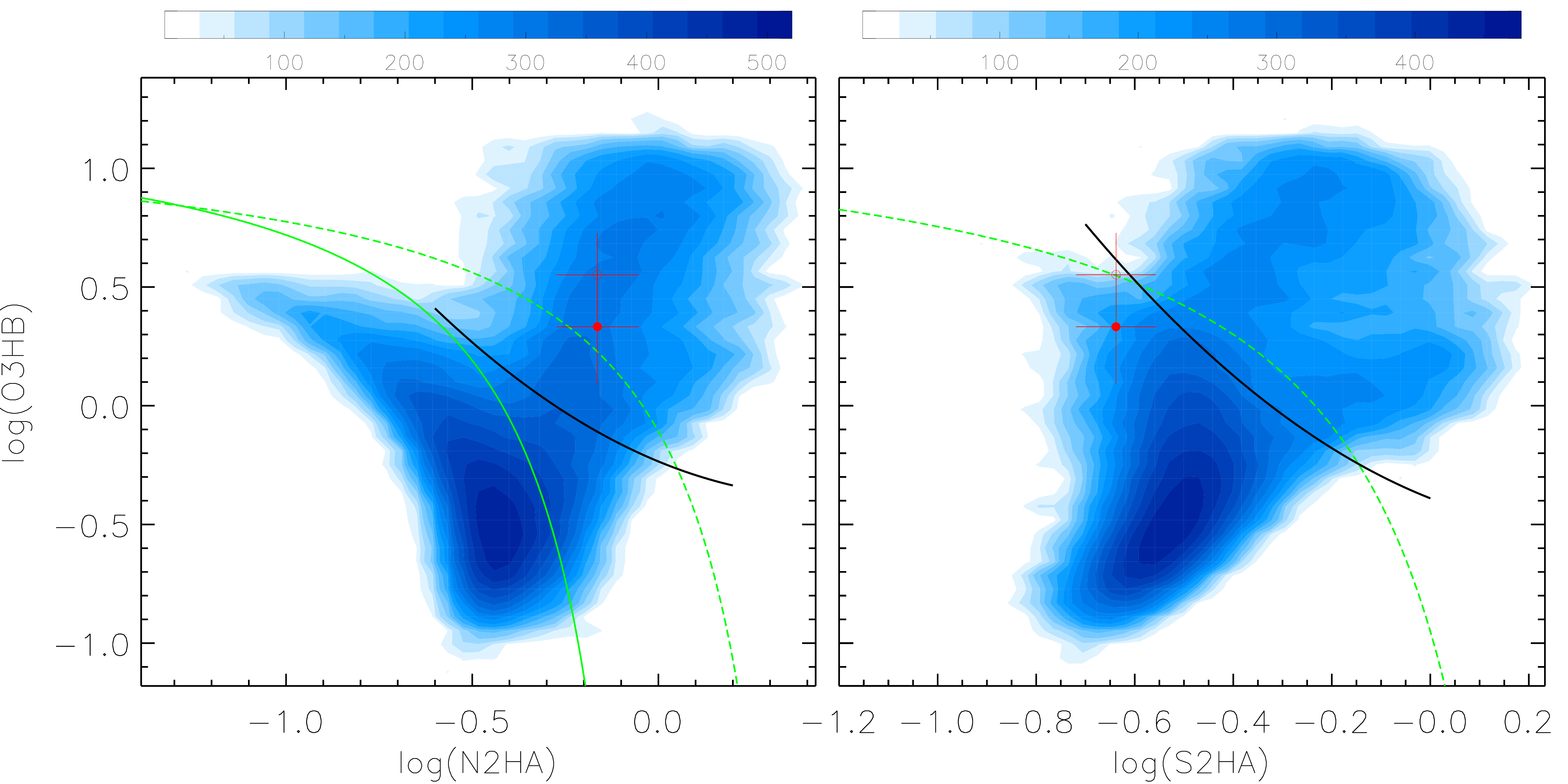}
\caption{Left panel shows the BPT diagram of O3HB versus N2HA, right panel shows the BPT diagram of O3HB 
versus S2HA. In each panel, contour in bluish colors shows the results for more than 35000 SDSS narrow line objects as 
described in \citet{zh20}, with corresponding number densities to different colors are shown in the color bar in top 
region of each panel. In each panel, solid (open) circle and error bars in red show the results of \obj~ without (with) 
considering contributions of the extended component in [O~{\sc iii}]$\lambda5007$\AA. In each panel, dashed green line 
shows the dividing line between AGN and HII galaxies reported in \citet{kb06}, solid black line shows the dividing line 
between AGN and HII galaxies reported in \citet{zh20}. In left panel, solid green line shows the dividing line between 
HII galaxies and composite galaxies reported in \citet{ka03b}.}
\label{bpt}
\end{figure*}

\begin{figure*}
\centering\includegraphics[width = 18cm,height=7.75cm]{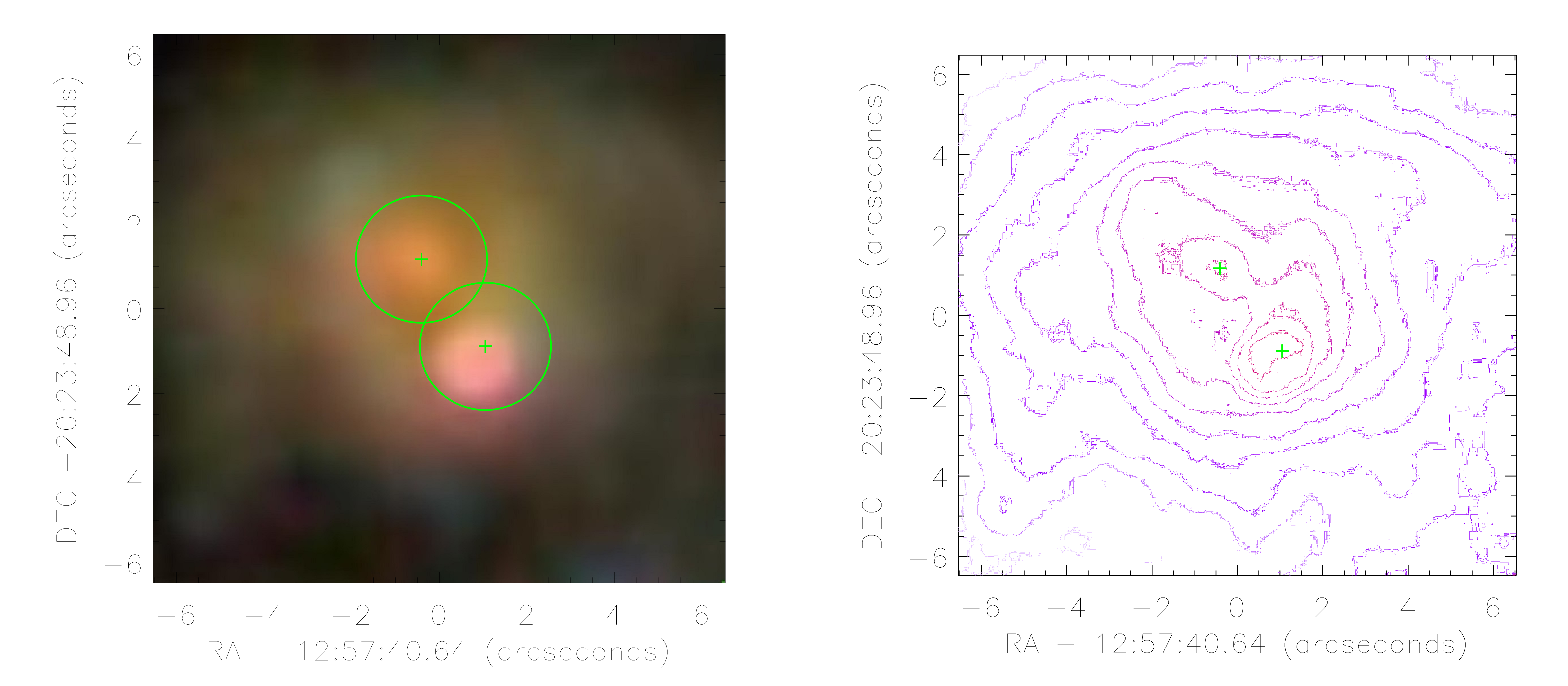}
\caption{Left panel shows the 13\arcsec$\times$13\arcsec~ colorful photometric image of \obj, right panel 
shows corresponding brightness contour of the photometric image. In left panel, the two green circles with radii 
1.5\arcsec~ represent covering regions of the SDSS fibers, the two green pluses represents the pointing positions 
(same as the central positions of the two nuclei after checking the inforamtion of 'plug\_ra' and 'plug\_dec' of 
the fibers for the SDSS spectra) of the SDSS fibers. In right panel, the two green pluses mark the central positions 
of the two nuclei, based on peak positions of the brightness contour.}
\label{img}
\end{figure*}

\begin{table}
\caption{Line parameters}
\begin{tabular}{lccc}
\hline\hline
	line & $\lambda_0$ & $\sigma$ & flux  \\
	&   \AA & \AA & ${\rm 10^{-17}erg/s/cm^2}$ \\
\hline\hline
Broad H$\alpha$ & 6565.41$\pm$0.42 & 24.11$\pm$0.92 & 1462.15$\pm$89.22 \\
\hline
Broad H$\beta$ & 4861.14$\pm$0.41 & 12.42$\pm$0.71 & 390.22$\pm$19.15 \\
\hline
\multirow{2}{*}{*broad H$\alpha$*} & 6562.53* & 16.77* & 711.23$\pm$289.14 \\
	& 6567.21$\pm$1.31 & 26.64$\pm$1.93 & 1013.56$\pm$182.16 \\
\hline\hline
\multirow{2}{*}{Narrow H$\alpha$}  &  6565.43$\pm$0.23  & 4.94$\pm$0.22  & 3713.25$\pm$88.43 \\
	& 6566.72$\pm$0.13 & 1.81$\pm$0.13 & 1256.56$\pm$120.44 \\
\hline
\multirow{2}{*}{Narrow H$\beta$}  &  4863.24$\pm$0.13  & 3.61$\pm$0.11  & 573.48$\pm$21.44 \\
	& 4864.31$\pm$0.11 & 1.31$\pm$0.11 & 216.23$\pm$19.21 \\
\hline
\multirow{4}{*}{[O~{\sc iii}]$\lambda5007$\AA} & 5008.61$\pm$1.82 & 2.62$\pm$0.72 & 618.35$\pm$490.16 \\
	& 5013.92$\pm$0.22 & 1.12$\pm$0.11 & 277.43$\pm$115.18 \\
	& 5011.41$\pm$0.22 & 1.73$\pm$0.31 & 803.55$\pm$504.37 \\
	& 5006.44$\pm$0.22 & 5.42$\pm$0.11 & 1115.23$\pm$74.46 \\
\hline
\multirow{4}{*}{[O~{\sc iii}]$\lambda4959$\AA} & 4960.71$\pm$1.79 & 2.62$\pm$0.71 & 213.33$\pm$167.14 \\
	& 4965.92$\pm$0.21 & 1.11$\pm$0.11 & 89.21$\pm$38.12 \\
	& 4963.42$\pm$0.21 & 1.71$\pm$0.31 & 270.34$\pm$171.36 \\
	& 4958.43$\pm$0.21 & 5.41$\pm$0.21 & 340.44$\pm$26.42 \\
\hline
\multirow{2}{*}{[N~{\sc ii}]$\lambda6583$\AA}  &  6584.71$\pm$0.21  & 4.93$\pm$0.22  & 1321.25$\pm$50.06 \\
	& 6587.51$\pm$0.13 & 2.23$\pm$0.13 & 1038.16$\pm$58.11 \\
\hline
\multirow{2}{*}{[N~{\sc ii}]$\lambda6548$\AA}  & 6549.51$\pm$0.21 & 4.93$\pm$0.21 & 451.33$\pm$404.08 \\
	& 6552.43$\pm$0.13 & 2.23$\pm$0.12 & 597.41$\pm$286.09 \\
\hline
\multirow{2}{*}{[S~{\sc ii}]$\lambda6716$\AA} & 6720.42$\pm$0.11 & 1.83$\pm$0.21 &  124.12$\pm$29.09 \\
	& 6719.52$\pm$0.32 & 4.31$\pm$0.21 & 478.23$\pm$37.08 \\
\hline
\multirow{2}{*}{[S~{\sc ii}]$\lambda6731$\AA} & 6734.92$\pm$0.11 & 1.90$\pm$0.31 & 168.38$\pm$59.19 \\
	& 6733.53$\pm$0.42 & 3.92$\pm$0.41 & 373.12$\pm$48.34 \\
\hline\hline
\end{tabular}\\
Notice: For the two rows related to '*broad H$\alpha$*', the first component is the one with the same redshift and 
the same second moment (in unit of ${\rm km/s}$) as the broad component in H$\beta$, the second component is the 
additional broad Gaussian component included in broad H$\alpha$.
\end{table}

\section{Spectroscopic properties of \obj}

	Based on one of our ongoing projects to determine whether dual core systems (at kpc scales) are preferred to 
explain double-peaked narrow emission lines motivated by our previous discussions in \citet{zh23}, there is a small 
sample\footnote{Detailed descriptions on the sample should be reported in one being prepared manuscript.} of tens of 
objects collected from SDSS, based on two cores well detected in photometric images and on spectroscopic results for 
the two cores. Then, after well measured broad emission lines, \obj~ is collected as the target of the manuscript, 
due to its quite different line widths of broad Balmer emission lines and then detected optical QPOs in its long-term 
variabilities, motivated by our previous results in \citet{zh21d}.

	Fig.~\ref{spec} shows the high-quality SDSS spectrum (signal-to-noise about 36, 2615-54483-0570 as 
PLATE-MJD-FIBERID) of \obj~ (RA=12:57:41.04,~DEC=+20:23:47.78, at redshift 0.081) collected from SDSS DR16 \citep{ap21}. 
\obj~ is classified as a quasar by the SDSS pipeline, however, due to apparent stellar features around 4000\AA, host 
galaxy contributions should be firstly subtracted before measuring emission lines.

	The widely accepted SSP (Simple Stellar Population) method has been well applied to determine host galaxy 
contributions in SDSS spectrum of \obj. Detailed discussions and descriptions on the SSP method can be found in 
\citet{bc93, bc03, ka03, cm05, cm17, wc19}. And the known SSP method has been well applied in our more recent papers 
in \citet{zh21a, zh21b, zh21c, zh21e, zh22b, zh22c}. Therefore, there are no detailed discussions on the SSP method, 
but simple descriptions are given as follows in the manuscript. The 39 simple stellar population templates in \citet{ka03} 
have been exploited, which can be used to well-describe the characteristics of almost all the SDSS galaxies as detailed 
discussions in \citet{ka03}. Meanwhile, there is an additional power law component, which is applied to describe intrinsic 
AGN continuum emissions. The SDSS spectrum with both broad and narrow emission lines being masked out can be well 
described by linear combinations of the broadened and shifted stellar templates plus a power law component, through 
the Levenberg-Marquardt least-squares minimization technique (the known MPFIT package). Left panels show best descriptions 
and corresponding line spectrum (SDSS spectrum minus the best descriptions) of \obj, leading $\chi^2/dof$ (the summed 
squared residuals divided by the degree of freedom) to be around 1.45. Meanwhile, stellar velocity dispersion 
can be well determined as 116$\pm$12km/s (the broadening velocity of the stellar templates), which will be simply applied 
to estimate central BH mass of \obj~ in the manuscript. Here, right panels of Fig.~\ref{spec} show detailed fitting 
results to absorption features around 4000\AA, to support the measured stellar velocity dispersion.

	After subtractions of SSP method determined host galaxy contributions, strong emission lines in the line 
spectrum can be measured by multiple Gaussian functions, similar as what we have more recently done in \citet{zh21b, 
zh21c, zh21d, zh22a, zh22b, zh22c, zh22d}. For broad and narrow H$\beta$ within rest wavelength range from 4800 to 
4900\AA, there are three Gaussian functions applied. For [O~{\sc iii}]$\lambda4959,5007$\AA~ doublet with rest 
wavelength from 4930 to 5040\AA, there are four Gaussian functions applied to each [O~{\sc iii}] line, in order 
to find the best descriptions. For H$\alpha$ and [N~{\sc ii}] doublet within rest wavelength from 6450 to 6670\AA, 
there are three Gaussian functions applied to describe broad and narrow H$\alpha$, and two Gaussian functions applied 
to each [N~{\sc ii}] line including a core component and a blue-shifted wing. For [S~{\sc ii}]$\lambda6716,6731$\AA~ 
doublet within rest wavelength from 6690 to 6760\AA, there are two Gaussian functions applied to each [S~{\sc ii}] 
line including a core component plus a blue-shifted wing. Moreover, besides Gaussian components for emission lines, 
power law components are applied to describe continuum emissions underneath the emission lines. When the Gaussian 
functions above are applied, each Gaussian component has flux not smaller than zero, and the corresponding Gaussian 
components in each [O~{\sc iii}] ([N~{\sc ii}], [S~{\sc ii}]) emission line have the same redshift. Then, through 
the Levenberg-Marquardt least-squares minimization technique, the best fitting results to the emission lines and 
corresponding residuals (line spectrum minus the best fitting results and then divided by uncertainties of SDSS 
spectrum) are shown in Fig.~\ref{ha} with determined $\chi^2/dof$ marked in title of each panel. The measured line 
parameters of each Gaussian component are listed in Table~1 with center wavelength $\lambda_0$ in unit of \AA, line 
width (second moment) $\sigma$ in unit of \AA~ and flux in unit of ${\rm 10^{-17}~erg/s/cm^2}$.

        Based on the measured line parameters, three points can be reported. First, considering line width (second 
moment) about 300${\rm km/s}$ of the broad components in [O~{\sc iii}], [N~{\sc ii}] and [S~{\sc ii}] doublets, the 
intermediate broad components with line width (second moment) about 220${\rm km/s}$ in H$\beta$ and in H$\alpha$ 
are accepted as components coming from narrow Balmer emission line regions. Second, in Fig.~\ref{bpt}, 
the \obj~ is shown in the BPT diagram of flux ratio O3HB of [O~{\sc iii}]$\lambda5007$\AA~ to narrow H$\beta$ versus 
flux ratio N2HA of [N~{\sc ii}]$\lambda6548,6583$\AA~ to narrow H$\alpha$, and in the BPT diagram of flux ratio of 
O3HB versus flux ratio S2HA of [S~{\sc ii}]$\lambda6716,6731$\AA~ to narrow H$\alpha$. In the BPT diagram of O3HB 
(with and without considerations of extended component in [O~{\sc iii}]$\lambda5007$\AA~ line) versus N2HA, the 
\obj~ lies above the dividing line between composite galaxy and AGN as well discussed in \citet{ka03b, kb06, zh20}, 
indicating apparent central AGN activities in \obj. Meanwhile, in the BPT diagram of O3HB versus S2HA, the \obj~ 
lies below the dividing line between HII galaxies and AGN, probably due to strong starforming contributions in \obj~ 
which have few effects on our properties of broad emission lines in \obj. Therefore, properties of \obj~ in the BPT 
diagrams and the determined power law continuum emission component in SDSS spectrum of \obj~ can be well applied to 
support that the broad Balmer emission lines are strongly related to central AGN activities of \obj, and also to 
probably support that AGN driven outflows should have contributions to the complicated [O~{\sc iii}] emission lines 
of \obj~ as well discussed in \citet{zh21b}. Third, the measured broad H$\alpha$ has line width (second moment) 
about 1100${\rm km/s}$ (full width at half maximum about 2600${\rm km/s}$), however the broad H$\beta$ has line width 
only about 760${\rm km/s}$ (FWHM about 1800${\rm km/s}$), indicating the broad H$\alpha$ are about 1.45 times wider 
than the broad H$\beta$ in \obj.

	Before proceeding further, necessary four points are noted. First, as reported in \citet{ls11}, there is a dual 
core system around \obj. The 13\arcsec$\times$13\arcsec~ colorful photometric image and corresponding 
brightness contour are shown in Fig.~\ref{img} around \obj~ \footnote{Photometric images with different scales around 
\obj~ can be found in\url{https://skyserver.sdss.org/dr16/en/tools/chart/navi.aspx?ra=194.421036563735&dec=20.3966075055986}}. 
Based on the shown photometric images in Fig.~\ref{img}, the projected space separation about 2.5\arcsec (about 4500pc) 
can be estimated of the central two cores. Furthermore, information of 'plug\_ra' and 'plug\_dec' has been carefully checked 
for the SDSS fiber for the spectrum of \obj, which is well consistent with the determined central position of \obj~ in the 
photometric image, therefore, it can be safely accepted that the SDSS fiber is correctly pointed to the center of \obj~ 
determined in the photometric image. Meanwhile, a small region is not only covered by the fiber for \obj~ but also 
covered by the fiber for the companion galaxy, indicating 
probable contributions from companion galaxy to the SDSS spectrum of \obj. Therefore, the galaxy merging process could 
lead to complicated narrow emission lines, and that is why multiple Gaussian functions rather than single Gaussian function 
are applied to describe each narrow emission line in \obj. However, based on the measured continuum luminosity about 
$1.97\times10^{43}{\rm erg/s}$ at 5100\AA~ through the determined power law continuum emissions shown in top left panel 
of Fig.~\ref{spec}, the estimated BLRs size (distance between BLRs and central black hole) in \obj~ is about 15 light-days 
(0.012pc) through the empirical dependence of BLRs size on continuum luminosity as well discussed in \citet{ks00, bd13}, 
clearly indicating that the dual core system has no effects on observed central broad emission lines from BLRs of \obj. 
To discuss properties of narrow emission lines are not objective of the manuscript, therefore, there are no further 
discussions on the dual core system or the complicated narrow emission lines.

	Second, besides the best fitting results shown in Fig.~\ref{ha} to emission lines in the line spectrum after 
subtractions of host galaxy contributions, similar fitting procedures are applied to describe emission lines in the 
SDSS spectrum without subtractions of host galaxy contributions in \obj, slightly different line parameters of 
narrow Balmer lines can be found, but totally similar results on broad H$\alpha$ and broad H$\beta$ can be confirmed, 
indicating host galaxy contributions have no effects on measurements of broad Balmer emission lines.

	Third, due to complicated line profiles of [O~{\sc iii}] doublet, less than four Gaussian functions have been 
applied to describe each [O~{\sc iii}] line, however, worse fitting results were obtained with $\chi^2/dof$ to be 3.37 
for three Gaussian functions applied and to be 6.73 for two Gaussian functions applied. Therefore, the four Gaussian 
functions are preferred to describe each [O~{\sc iii}] line, although there are large determined uncertainties of 
fluxes of the components. Therefore, there are no further discussions in the manuscript on the fitting results to 
[O~{\sc iii}] emission lines by different number of Gaussian functions.

%Here, Fig.~\ref{o32} 
%shows fitting results and corresponding residuals with larger scatters to [O~{\sc iii}] doublet by 
%two or three Gaussian functions applied, but there are no further discussions on the results in 
%Fig.~\ref{o32} in the manuscript. 

	Fourth, in order to find the best descriptions to emission lines, flux is a free parameter in each Gaussian 
component in [O~{\sc iii}] and [N~{\sc ii}] doublets. Therefore, parameters of Gaussian components in 
[O~{\sc iii}]$\lambda4959$\AA~ and in [N~{\sc ii}]$\lambda6548$\AA~ are also listed in Table~1. And measured flux 
ratios of corresponding Gaussian components in [O~{\sc iii}] ([N~{\sc ii}]) doublet are well around the theoretical 
value 3 after considering measured flux uncertainties, as discussed in \citet{dp07, dk22}.

	Furthermore, in order to show clearer difference between broad H$\alpha$ and broad H$\beta$, besides three 
Gaussian functions applied to describe H$\alpha$ shown in bottom left panel of Fig.~\ref{ha}, the line profile of 
H$\beta$ is applied to describe H$\alpha$ by three Gaussian components having the same redshift and the same second 
moment (in velocity space) as those of H$\beta$, leading the $\chi^2/dof$ to be about 4.2, even considering three 
Gaussian components included in each [N~{\sc ii}] emission line. The fitting results and corresponding residuals 
with larger scatters are shown in top panel of Fig.~\ref{ha2}. Worse descriptions to the wings (around 6520\AA~ and 
6610\AA, marked by vertical dashed cyan lines) around the emission lines can be found, indicating broader components 
are necessary. Meanwhile, in bottom panel of Fig.~\ref{ha2}, the line profile of H$\beta$ plus an addition Gaussian 
component are applied to describe H$\alpha$, leading $\chi^2/dof$ to be about 1.08, indicating the additional Gaussian 
component is preferred. Parameters of the additional broad Gaussian component are also listed in Table~1. Therefore, 
there are quite different line widths of broad H$\alpha$ and broad H$\beta$. Moreover, considering the quite different 
values of $\chi^2/dof$ through applications of different model functions, the additional broad Gaussian component 
can be well confirmed and preferred with confidence level well higher than 5$\sigma$ through applications of F-test 
statistical technique as what we have recently done in \citet{zh22d}. In other words, the quite different line widths 
of broad Balmer lines can be confirmed with confidence level higher than 5$\sigma$ in \obj.

\begin{figure}
\centering\includegraphics[width = 8cm,height=5cm]{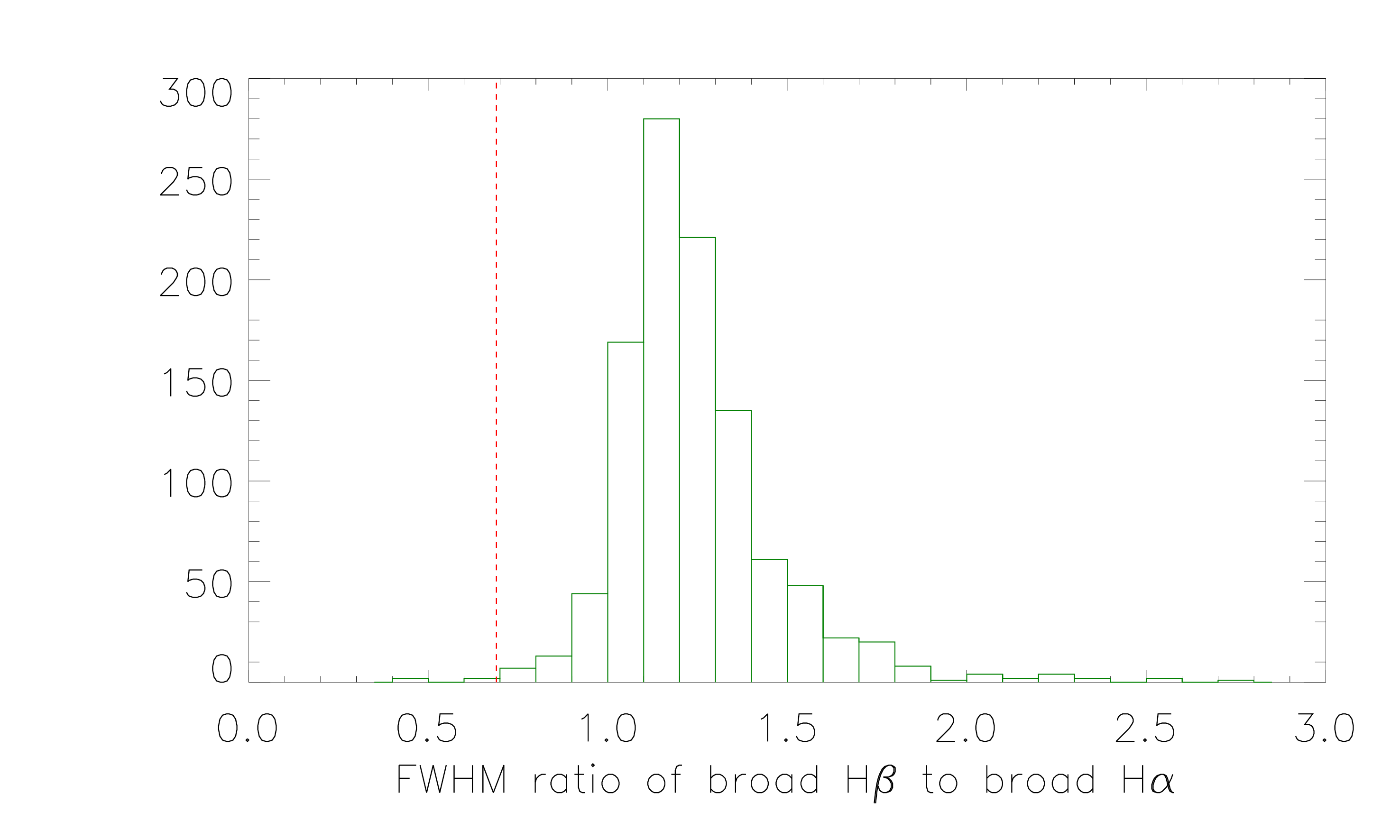}
\caption{Distribution of FWHM ratio of broad H$\beta$ to broad H$\alpha$ of the collected 1048 normal 
QSOs from \citet{sh11}. Vertical dashed red line marks the position for the FWHM ratio to be 0.69 (the ratio in \obj).	
}
\label{unu}
\end{figure}

	As well discussed in \citet{gh05}, there is a strong linear FWHM (full width at half maximum, 2.35 times 
of the second moment of Gaussian-like broad lines) correlation between broad H$\alpha$ and broad H$\beta$ of a large 
sample of SDSS quasars, and the FWHM ratio is about 1.1 (scatter about 0.1dex) of broad H$\beta$ to broad H$\alpha$. 
Meanwhile, based on the reported measurements of FWHMs of broad H$\beta$ and broad H$\alpha$ of 1048 
SDSS QSOs in \citet{sh11} with redshift smaller than 0.3 and with reliable measurements of FWHMs (at least 10 times 
larger than their uncertainties), Fig.~\ref{unu} shows the distribution of FWHM ratio of broad H$\beta$ to broad 
H$\alpha$ of the collected 1048 SDSS quasars from \citet{sh11}. Among the collected 1048 SDSS QSOs, there are only 
4 QSOs with FWHM ratio smaller than 0.69. Therefore, the \obj~ with line width ratio about 0.69$\pm$0.06 of broad 
H$\beta$ to broad H$\alpha$ strongly indicates that \obj~ is an unique quasar among SDSS quasars.

%\iffalse
%Moreover, based on the reported line parameters of 551 SDSS quasars with redshift less than 
%0.3\ in \citet{sh11} with reliable broad Balmer emission lines of which both FWHMs and line luminosities are at least 
%10 times larger than their corresponding uncertainties and of which signal-to-noise around Balmer lines are larger than 
%20, distribution of FWHM ratio of broad H$\beta$ to broad H$\alpha$ is shown in Fig.~\ref{dis}, and there are only three 
%quasars (which will be discussed in the following subsection) with the ratio smaller than 0.69, to re-confirm that the 
%\obj~ is an unique object.
%\fi

	In order to explain broad H$\alpha$ wider than broad H$\beta$, the simplest case on two independent BLRs 
related to a BBH system can be considered. For a BBH system expected two broad Balmer lines from two independent 
BLRs related to two central active nuclei, each observed broad Balmer line can be described by sum of two broad Balmer 
line emission components. Considering simple results shown in bottom panel of Fig.~\ref{ha2}, the broad component with 
larger second moment being seriously obscured can naturally lead to quite different line widths of broad H$\alpha$ 
and broad H$\beta$, simply as what we have discussed in \citet{zh21d}. Meanwhile, detailed discussions 
in the following Section 4 should provide further clues to support different obscurations on central two BLRs related 
to a BBH system can lead to different line profiles of broad Balmer emission lines in AGN. In order to support the 
assumed BBH system, it is interesting to check long-term variabilities of \obj.

\section{Long-term variabilities of \obj}

\begin{figure}
\centering\includegraphics[width = 8cm,height=11cm]{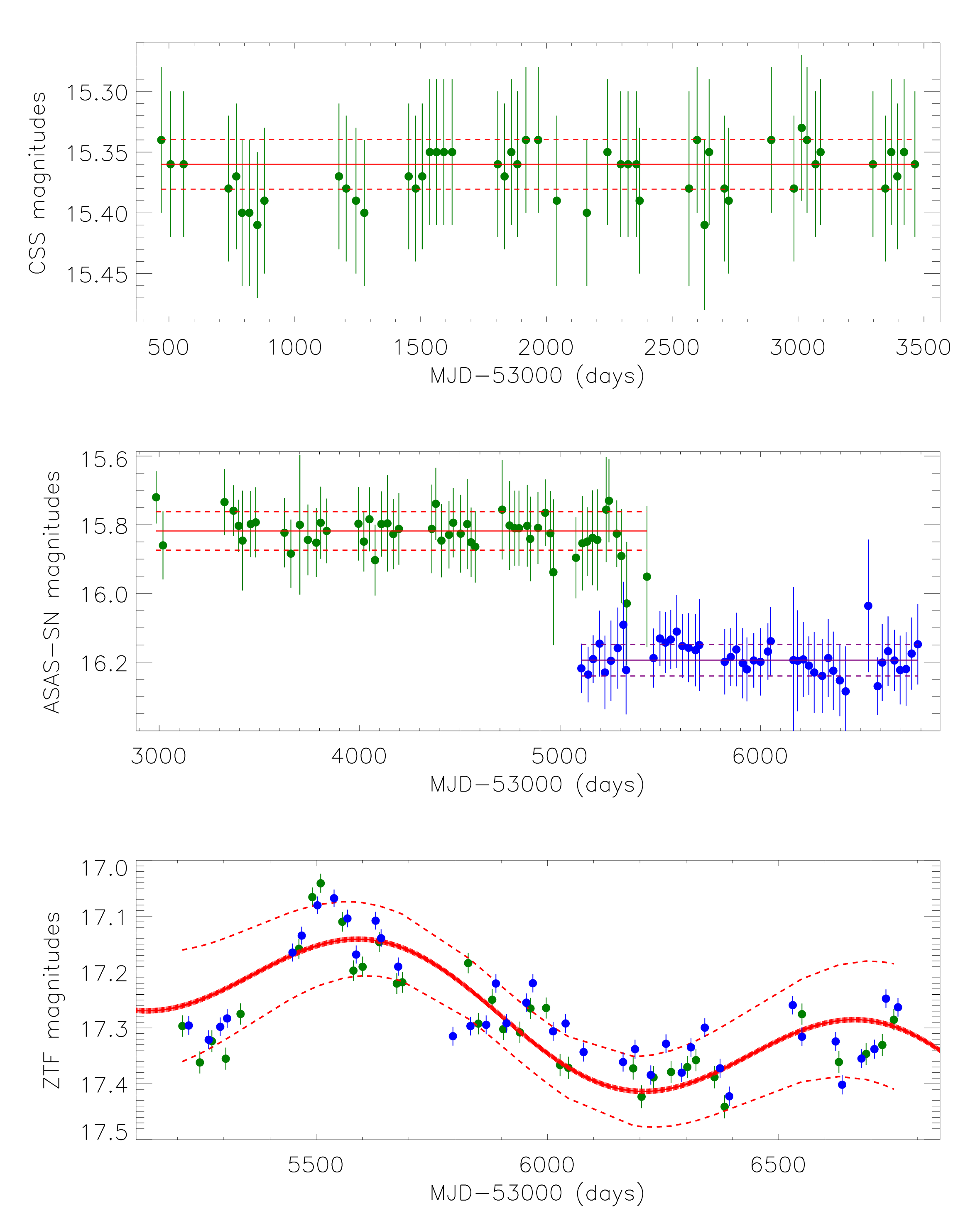}
\caption{The long-term 30days-binned light curves from the CSS (solid circles plus error bars in dark green) in top 
panel, and from the ASAS-SN (dark green for V-band, blue for g-band) in middle panel, and from the ZTF (dark green 
for g-band, and blue for r-band plus 0.67 magnitudes) in bottom panel. In top panel and middle panel, solid and 
dashed lines show the mean value and corresponding 1RMS scatter bands of the light curve. In bottom panel, solid 
red line and dashed red lines show the best fitting results by a sinusoidal function plus a linear trend and 
corresponding $3\sigma$ confidence bands to the ZTF light curves.}
\label{lmc}
\end{figure}

	As describes in the Introduction, optical QPOs are best indicators of BBH systems. Therefore, 
besides the different line profiles of broad Balmer emission lines in \obj, it is necessary to check whether 
optical QPOs can be detected in long-term variabilities of \obj, which will provide further evidence to support 
a BBH system in central region of \obj, and then provide strong evidence to support that different line profiles 
of broad Balmer lines can be accepted as an indicator of central BBH systems in AGN.

	The CSS V-band light curve of \obj~ is collected from \url{http://nesssi.cacr.caltech.edu/DataRelease/} 
\citep{dr09} with MJD-53000 from 469 (Apr. 2005) to 3464 (Jun. 2013). And the ASAS-SN (All-Sky Automated Survey 
for Supernovae) \citep{sp14, ko17} V/g-band light curves are collected with MJD-53000 from 2985 (Feb. 2012) to 
6795 (Aug. 2022). However, there are no apparent variabilities in the light curves from CSS or from ASAS-SN, 
probably due to larger magnitude uncertainties relative to intrinsic variabilities. Therefore, there are no 
further discussions on the light curves from CSS or ASAS-SN, but the light curves are shown in top panel and 
middle panel of Fig.~\ref{lmc} with clearly marked 1RMS scatter bands.

\begin{figure}
\centering\includegraphics[width = 8cm,height=10cm]{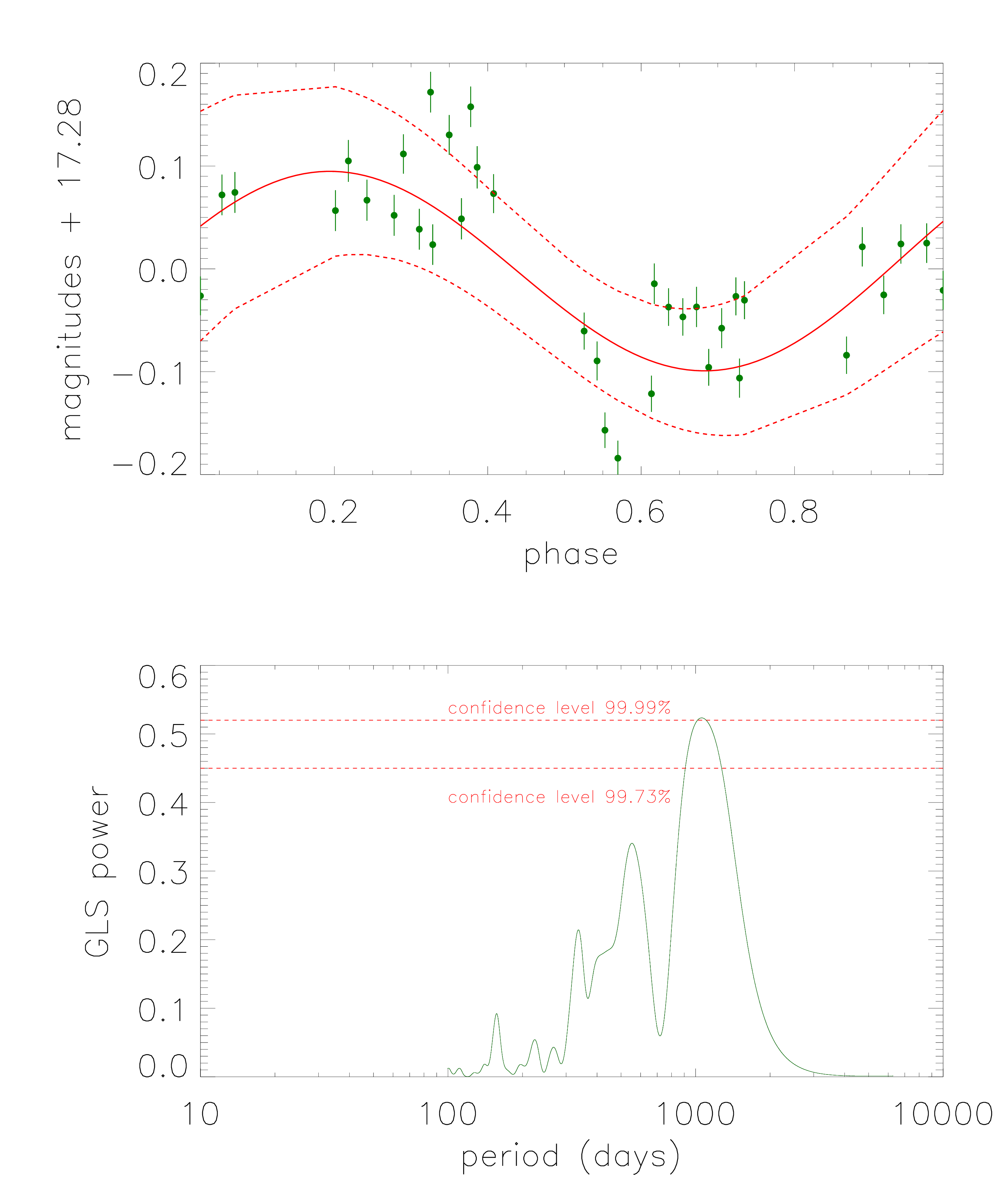}
\caption{Top panel shows the phase-folded light curve of \obj, accepted periodicity about 1070days. Solid red line 
and dashed red lines show the best fitting results by the simple sinusoidal function plus a linear trend and 
corresponding $3\sigma$ confidence bands. Bottom panel shows properties of GLS power of \obj. Horizontal red 
lines show positions relative to confidence levels of 99.99\% (false-alarm probability of 1e-4) and 99.73\% 
(false-alarm probability of 2.7$\times$1e-3).
}
\label{qpo}
\end{figure}

\begin{figure}
\centering\includegraphics[width = 8cm,height=10cm]{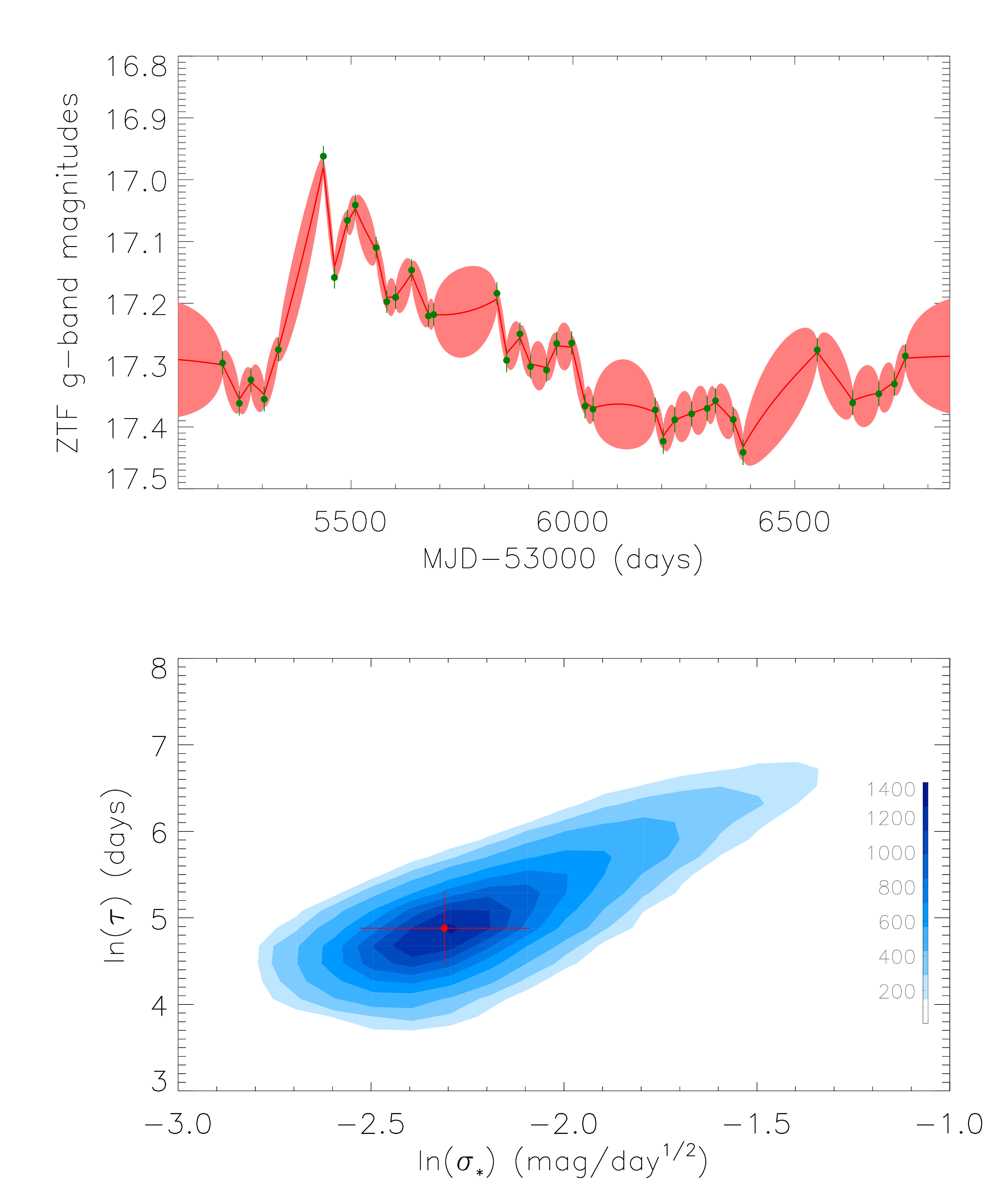}
\caption{Top panel shows the DRW-determined best descriptions to the long-term ZTF variabilities of \obj. Solid 
red line and area filled with light red show the best descriptions and corresponding 1sigma confidence bands. 
Bottom panel shows the MCMC determined two-dimensional posterior distributions of the DRW process parameters of 
$\ln(\tau)$ and $\ln(\sigma_*)$. In bottom panel, solid circle plus error bars in red show the accepted values 
and corresponding uncertainties of the two parameters, color bar shows the corresponding number densities related 
to different colors.
}
\label{drw}
\end{figure}

	Meanwhile, the ZTF g/r-band light curves of \obj~ are collected from \url{https://www.ztf.caltech.edu} 
with MJD-53000 from 5202 (Mar. 2018) to 6770 (Jul. 2022), with apparent variabilities shown in bottom panel of 
Fig.~\ref{lmc}. Then, through Levenberg-Marquardt least-squares minimization technique, a simple sinusoidal function 
plus a linear trend can be applied to well describe the ZTF light curves. Here, the main objective of application 
of sinusoidal function is to show clearer clues on optical QPOs, not to discuss physical origin of the QPOs. The 
best-fitting results with $\chi^2/dof~\sim~12.7$ (the summed squared residuals divided by degree of freedom) and 
corresponding $3\sigma$ confidence bands through F-test technique are shown as solid and dashed red lines in bottom 
panel of Fig.~\ref{lmc} by the formula
\begin{equation}
LMC~=~A~+~B~\times~\frac{t}{\rm 1000days}~+~C~\times~\sin(\frac{2\pi t}{T_{p}}~+~\phi_0)
\end{equation}
with $A=16.485\pm0.063$, $B=0.13\pm0.01$, $C=0.098\pm0.004$, $T_{p}=1070\pm30{\rm days}$, $\phi_0=3.198\pm1.014$, 
leading to QPOs with a periodicity about $1070\pm30$ days. The sinusoidal variability pattern can be accepted as 
interesting clues to support optical QPOs in \obj.

	Although, the collected ZTF light curves have short time durations (1568days) only 1.47 times longer than 
the determined periodicity by sinusoidal function, it is interesting to re-confirm the optical QPOs through the 
other techniques. Similar as what we have recently done in \citet{zh22a}, properties of the phase-folded light 
curve accepted the periodicity about 1070days are shown in top panel of Fig.~\ref{qpo}, which can also be well 
described by a sinusoidal function. Meanwhile, the generalized Lomb-Scargle (GLS) periodogram \citep{ln76, sj82, 
zk09, vj18} is applied to check the optical QPOs, properties of the GLS power are shown in bottom panel of 
Fig.~\ref{qpo}. Through confidence level (related to false-alarm probability applied in the GLS) determined by 
the bootstrap method as discussed in \citep{ic19}, confidence level higher than $3\sigma$ (99.73\%) can be 
confirmed to support the optical QPOs with periodicity around 1000days. Therefore, although short time durations 
of the collected ZTF light curves, the optical QPOs can be accepted with confidence level higher than $3\sigma$ 
through the GLS periodogram technique.

	Before giving a robust conclusion on the optical QPOs in \obj, the following point should be mainly considered. 
As well known that intrinsic long-term variabilities are fundamental characteristics of AGN \citep{ww95, um97, kh04, 
sm10, mg15, bg18, bs21, yd22}, it is interesting and necessary to check whether the detected optical QPOs in \obj~ are 
actually related to intrinsic AGN activities. Therefore, the following method is applied, as what we have more 
recently done in \citet{zh22a}. The commonly known CAR (Continuous AutoRegressive) process can be well applied 
to check whether the QPOs signs in \obj~ shown in bottom panel of Fig.~\ref{lmc} were generally from intrinsic AGN 
activities which can be well modeled by CAR process (or the improved damped random walk process (DRW process)) as 
discussed in \citet{kbs09, koz10, mi10, zk13, kb14, sh16, zk16, zh17, tm18, mv19, bs21, sr22}. Here, the long-term 
ZTF variabilities of \obj~ can be firstly described by the public code JAVELIN (Just Another Vehicle for Estimating 
Lags In Nuclei) \citep{koz10, zk13} with two DRW process parameters of intrinsic characteristic variability amplitude 
and timescale of $\sigma_*$ and $\tau$. The best descriptions to the 30days-binned g-band light curve are shown in 
top panel of Fig.~\ref{drw}. And bottom panel of Fig.~\ref{drw} shows the corresponding MCMC (Markov Chain Monte 
Carlo) \citep{fh13} determined two dimensional posterior distributions of the process parameters of $\sigma_*$ 
and $\tau$, with the determined $\ln(\tau/{\rm days})\sim4.88\pm0.43$ ($\tau\sim132$days) and 
$\ln(\sigma_*/({\rm mag/days^{1/2}}))\sim-2.31\pm0.22$ ($\sigma_*\sim0.099{\rm mag/day^{1/2}}$).

	Then, probability of mis-detected QPOs from DRW process described intrinsic AGN variabilities can be 
simply estimated as follows. Based on the CAR process discussed in 
\citet{kbs09}: 
\begin{equation}
\dif LMC_t~=~\frac{-1}{\tau}LMC_t\dif t~+~\sigma_c\sqrt{\dif t}\epsilon(t)~+~17.28
\end{equation}
where $\epsilon(t)$ is a white noise process with zero mean and variance equal to 1, it is interesting 
to check whether can CAR process determined long-term variability lead to mis-detected QPOs. Here, the mean 
value of $LMC_t$ is set to be 17.28 (the mean value of the ZTF g-band light curve of \obj). Then, a series 
of 100000 simulating light curves [$t$,~$LMC_t$] are created, with $\tau$ set to be 132 days (the $\tau$ 
value of \obj) and $\tau\sigma_c^2/2$ set to be 0.012 (the variance of the ZTF light curve of \obj) (the 
parameter $\sigma_c$ in unit of mag in the CAR process in \citet{kbs09} slightly different from the JAVELIN 
determined $\sigma_*$). And, time information $t$ are the same as the observational g-band time information 
shown in bottom panel of Fig.~\ref{lmc}. And the similar uncertainties $LMC_{t,~err}$ are simply added to 
the simulating light curves $LMC_t$ by
\begin{equation}
LMC_{t,~err}~=~LMC_{t}\times\frac{L_{err}}{L_{obs}}
\end{equation}
with $L_{obs}$ and $L_{err}$ as the observational ZTF light curve and the corresponding uncertainties shown 
in bottom panel of Fig.~\ref{lmc}.

\begin{figure}
\centering\includegraphics[width = 8cm,height=16cm]{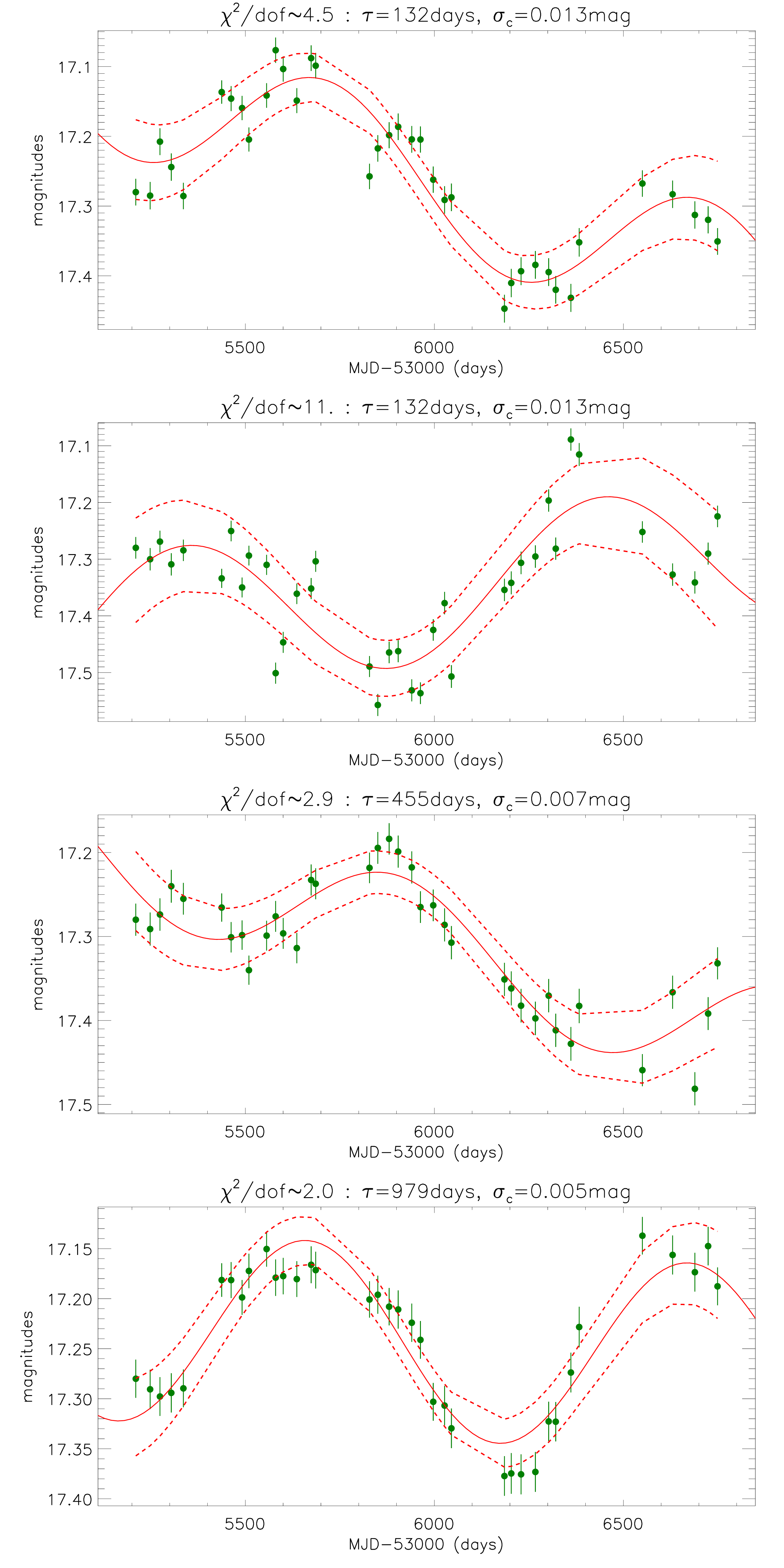}
\caption{On probable mis-detected QPOs in the CAR process simulating light curves with fixed $\tau$ in top two 
panels and with randomly selected $\tau$ in bottom two panels. In each panel, solid circles plus error bars in 
dark green show the CAR process simulated light curve, solid red line and dashed red lines show the best fitting 
results by a sinusoidal function plus a linear trend and corresponding $3\sigma$ confidence bands to the simulated 
light curve. The applied $\tau$, $\sigma_c$ and determined $\chi^2/dof$ are marked in title of each panel.
}
\label{fake}
\end{figure}

        Then, the following two simple criteria are applied to determine probable QPOs detected in the simulating 
light curves. First, one simulating light curve can be best described by the equation (1) with $\chi^2/dof<15$ 
($\chi^2/dof\sim12.7$ for the results shown in bottom panel of Fig.~\ref{lmc}), and the best-fitting procedure 
determined periodicity is between [1070-90days,~1070+90days] (90days as 3times of 30days determined periodicity 
uncertainty in bottom panel of Fig.~\ref{lmc}). Second, confidence level is higher than 99.73\% for the GLS power 
determined peak position. Finally, among the 100000 simulating light curves, there are 1027 light curves with 
expected mis-detected QPOs with periodicity around 1070days, accepted the two criteria above. Moreover, top two 
panels of Fig.~\ref{fake} shows 2 of the 1027 simulating light curves with mis-detected QPOs and the corresponding 
best-fitting results by equation (1) and F-test technique determined $3\sigma$ confidence bands. The results indicate 
that the CAR process related to central AGN activities can lead to light curves with mathematical determined QPOs 
(the mis-detected QPOs, or the fake QPOs), however, the probability of the mis-detected QPOs in CAR-process 
simulating light curves is around 1.03\% (1027/100000). The results strongly indicate that the probability 
higher than 98.97\% (1-1.03\%) (confidence level higher than 2.5$\sigma$) to support that the detected optical 
QPOs in \obj~ are not mis-detected QPOs from a pure CAR process described light curve, although the collected 
ZTF light curves have short time durations.

	Furthermore, when the simulating $LMC_t$ above are created, the parameter $\tau$ is fixed. If the parameter 
$\tau$ is randomly collected, are there different results? Then, a new series $LMC_t$ are created as follows. The 
DRW process parameter $\tau$ is randomly selected from 50days to 1000 days, common values for SDSS quasars discussed 
in \citet{mi10}. Then, another 100000 light curves are created by the CAR process. And then, based on the same 
criteria above, among the 100000 simulating light curves, there are 4433 light curves with expected mis-detected 
QPOs. Bottom two panels of Fig.~\ref{fake} shows 2 of the 4433 simulating light curves with mis-detected QPOs and 
the corresponding best-fitting results by equation (1) and F-test technique determined $3\sigma$ confidence bands. 
The results strongly indicate that different input parameters of $\tau$ and $\sigma_c$ have tiny effects on the 
final results, and that the probability higher than 4.4\% (confidence level higher than 2$\sigma$) to support that 
the detected optical QPOs in \obj~ are not mis-detected QPOs from a pure CAR process described variabilities related 
to central AGN activities.

	Therefore, although short time durations in ZTF light curves, through CAR process simulated light curves to
trace central AGN activities, it can be confirmed with confidence level higher than 2$\sigma$ that the optical QPOs
are not from intrinsic AGN activities in \obj.

\begin{figure*}
\centering\includegraphics[width = 18cm,height=6cm]{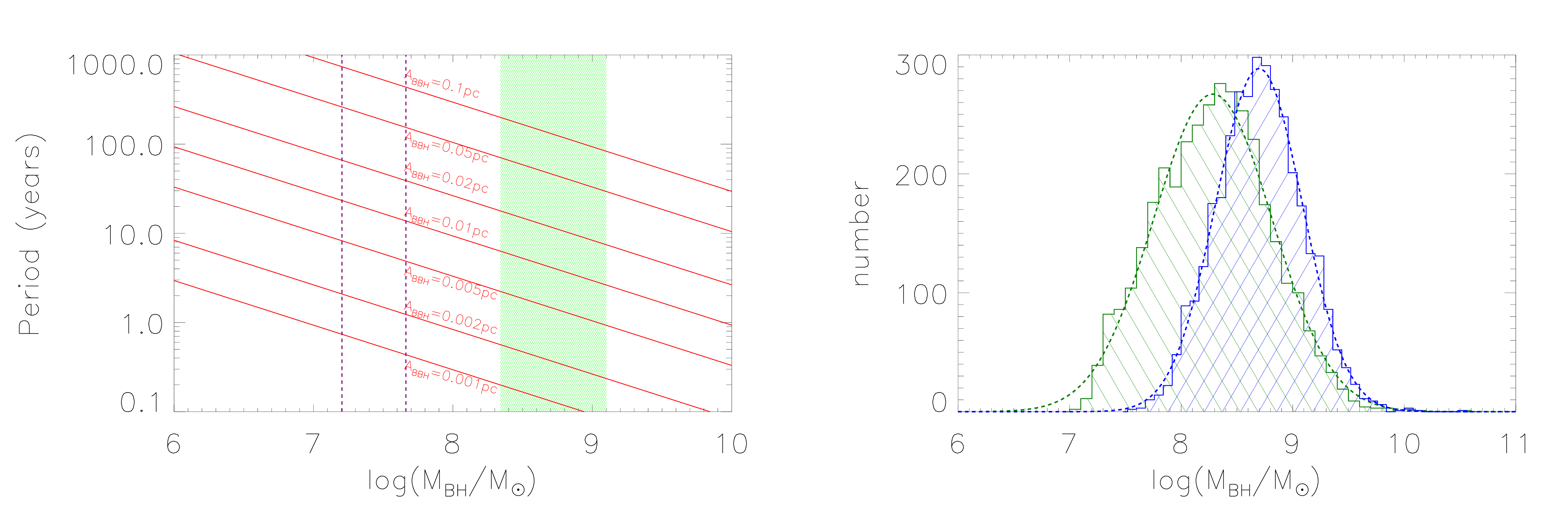}
\caption{Left panel shows dependence of expected periodicity on central total BH masses for given space 
separations of assumed BBH systems. Vertical dashed purple lines show lower and upper limits of total BH masses of 
\obj~ estimated by M-sigma relation, area filled by green lines marks ranges of total BH masses of BBH systems simply 
determined by BH mass distributions of local quasars reported in \citet{sh11}. Right panel shows BH mass distributions. 
Histogram filled by dark green lines show the BH mass distributions of local SDSS quasars in \citet{sh11}, dashed 
dark green line shows corresponding Gaussian description to the histogram. Histogram filled by blue lines show the 
total BH mass distributions of assumed BBH systems, dashed blue line shows corresponding Gaussian description to the 
histogram.}
\label{test_qpo}
\end{figure*}

	Before the end of the section, three points on simple properties of optical QPOs related to BBH systems can 
be discussed as follows. First and foremost, ideally maximum change in magnitudes (two times of the amplitude) of 
optical light curves related to BBH system determined QPOs can be simply determined to be about 0.75magnitudes 
($2.5\times\log(2)$). However, considering different inclination angles of BBH systems having different magnitudes 
of each BH accreting system, change in magnitudes of QPOs expected optical light curves should be smaller than 
0.75magnitudes, which can be confirmed by the reported light curves of optical QPOs in the literature, especially 
the reported results in \citet{gd15b} with changes about 0.1-0.6magnitudes in light curves of a sample of 111 
optical QPOs. Therefore, comparing with the shown results in reported QPOs, the change about 0.3magnitudes in light 
curve of the \obj~ in the manuscript can be well accepted.

	Besides, considering dependence of periodicity $P_{BBH}$ of BBH system expected optical QPOs on parameters 
of both total BH masses $M_{BH}$ and space separation $A_{BBH}$ of central two BH accreting systems as well discussed in 
\citet{eb12}
\begin{equation}
	\frac{A_{BBH}}{\rm pc}~=~0.432~(\frac{M_{BH}}{\rm 10^8M_\odot})^{1/3}~(\frac{P_{BBH}}{\rm 2652years})^{2/3}
\end{equation}
Left panel of Fig.~\ref{test_qpo} shows the dependence of $P_{BBH}$ on $M_{BH8}$ (in units of ${\rm 10^8M_\odot}$), 
for given $A_{BBH}$, totally similar as the shown results in \citet{gd15a, gd15b}. Then, two simple methods are applied 
to determined properties of total BH masses. On the one hand, based on reported BH masses of low redshift ($z<0.35$) 
quasars in \citet{sh11}, total BH mass properties of BBH systems can be simply determined as follows. Right panel of 
Fig.~\ref{test_qpo} shows BH mass distributions (histogram filled by dark green lines) of the 3501 low redshift quasars 
with reliable virial BH masses (at least 5 times larger than their corresponding uncertainties). Then, as a toy model, 
among the BH masses of the 3501 low redshift quasars, sum of two randomly selected BH masses can be roughly accepted 
as total BH mass of an assumed BBH system. Total BH masses of 3500 assumed BBH systems are created by sum of randomly 
selected BH masses, and shown as histogram filled by blue lines in the right panel of Fig.~\ref{test_qpo} with mean 
total BH mass $\log(M_{BH}/{\rm M_\odot})$ about 8.70 with standard deviation about 0.38. The total BH mass 
($\log(M_{BH}/{\rm M_\odot})$) range $8.70\pm0.38$ is shown as area filled by green lines in left panel of 
Fig.~\ref{test_qpo}. On the other hand, based on the M-sigma relation as well discussed in \citet{fm00, ge00, kh13, bb17, 
bt21}, total BH mass properties of \obj~ can be simply estimated as follows. Although a BBH system is expected in \obj, 
central total BH mass could be also estimated by the M-sigma relation, as well discussed in \citet{ds05, jn09}. Therefore, 
based on the measured stellar velocity dispersion about 116$\pm$16km/s, total BH mass of \obj~ (even a BBH system in 
central region) can be simply estimated to be $M_{BH}\sim2.8_{-1.2}^{+1.8}\times10^7{\rm M_\odot}$ through the M-sigma 
relation in \citet{kh13}. The M-sigma relation determined total BH mass with uncertainties are also shown in left panel 
of Fig.~\ref{test_qpo} for \obj. Based on the shown results in Fig.~\ref{test_qpo}, either on total BH masses of a 
sample of BBH systems or on total BH mass of \obj, it is clear that current detected timescales of optical QPOs related 
to BBH systems are around several years, after considering general time durations of current public sky survey projects 
and also restriction that time duration should be at least 1.5times longer than periodicities of optical QPOs.

	Last but not the least, as detailed discussions in \citet{gd15b} that periodic behavior is not expected as a 
result of a CAR process, and also as discussed above on probability of fake optical QPOs detected in long-term intrinsic 
AGN variabilities described by CAR process in the manuscript, optical QPOs have quite different properties from those of 
intrinsic AGN variabilities. Meanwhile, time durations longer enough in the near future could provide more robust 
evidence to support the optical QPOs in \obj.

\section{Further Applications}

	Accepted BBH system assumptions, simple results on line width (second moment) ratio of broad H$\beta$ to 
broad H$\alpha$ can be determined with considering different obscurations on central two independent BLRs, which 
will provide further clues to discuss different line properties of broad Balmer emission lines in near future. 
In the section, an oversimplified but interesting model can be considered as follows.

	One blue-shifted broad Gaussian component in broad H$\alpha$ related to one BH accreting system in one 
BBH system has central wavelength ($\lambda_{0b\alpha}$) randomly selected from 6564-60\AA~ to 6564.61\AA, and 
second moment ($\sigma_{0b\alpha}$) randomly selected from 15\AA~ to 90\AA, and line flux ($f_{0b\alpha}$ in 
arbitrary unit) randomly selected from 0.2 to 1. And the other red-shifted broad Gaussian component in broad 
H$\alpha$ related to the other BH accreting system in the BBH system has central wavelength ($\lambda_{0r\alpha}$) 
randomly selected from 6564\AA~ to 6564+60\AA, and second moment ($\sigma_{0r\alpha}$) randomly selected from 
15\AA~ to 90\AA, and line flux ($f_{0r\alpha}$) randomly selected from 0.2 to 1. Then, the simulating line profile 
of broad H$\alpha$ related to BBH system can be described by
\begin{equation}
f_\lambda(H\alpha)=G(\lambda,~[\lambda_{0b\alpha},~\sigma_{0b\alpha},
	~f_{0b\alpha}])+G(\lambda,~[\lambda_{0r\alpha},~\sigma_{0r\alpha}, ~f_{0r\alpha}])
\end{equation}
where $G(\lambda,~[\lambda_0,~\sigma_0, f_0])$ means a Gaussian emission component, and $\lambda$ means the rest 
wavelength from 6000\AA~ to 7000\AA~ for the simulated broad H$\alpha$. Actually, the wavelength range, second 
moment range and line flux range have tiny effects on our final results.

\begin{figure*}
\centering\includegraphics[width = 18cm,height=13cm]{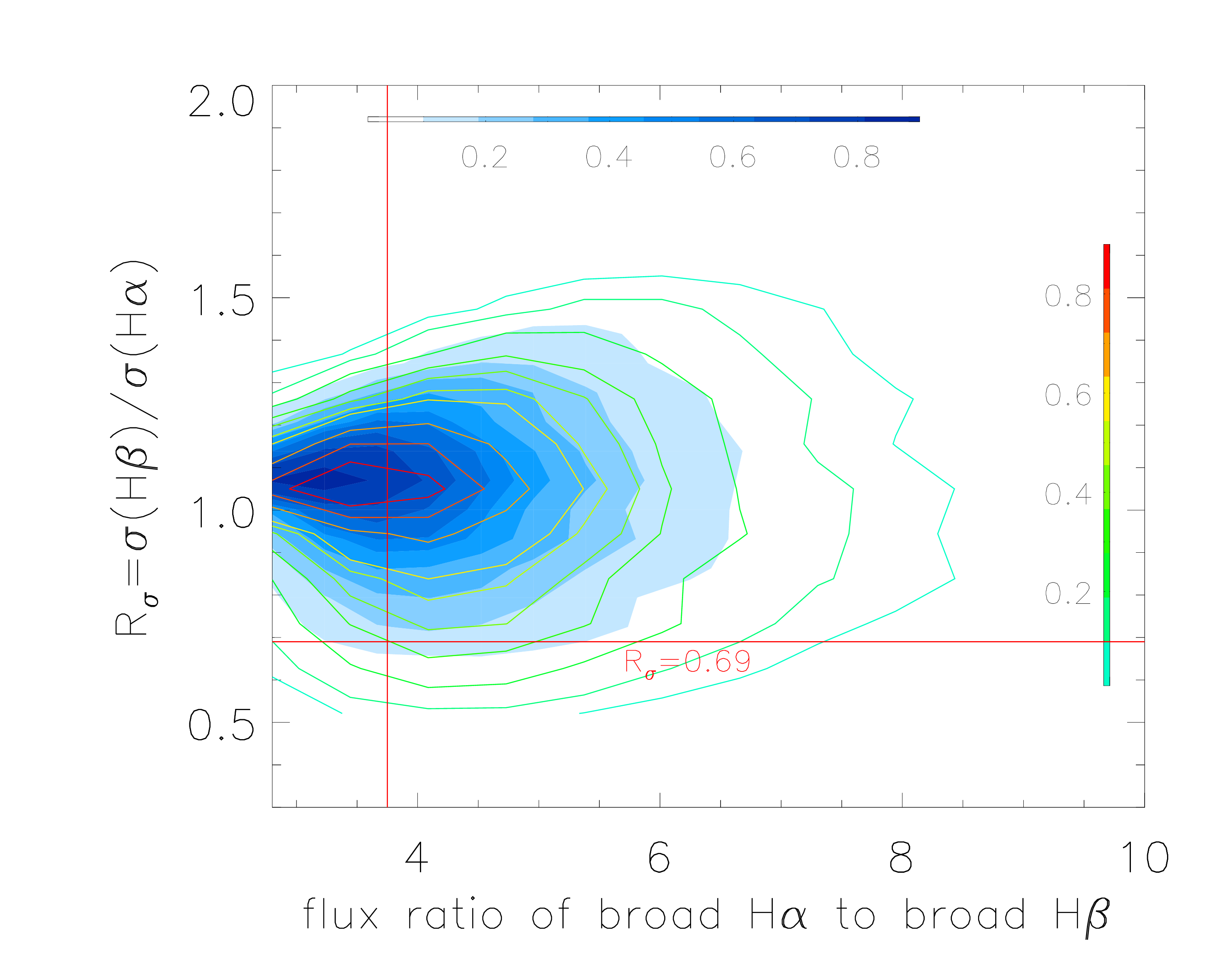}
\caption{Diagram of $R_\sigma$ versus flux ratio of simulated broad H$\alpha$ to simulated broad H$\beta$ shown 
in color filled contour for simulated results with normal distribution of $BD$ and shown in contour with levels 
shown in lines in different colors for simulated results with uniform distribution of $BD$. Contour 
levels with different colors represent where 5\% to 90\% of the two-dimensional volume are contained, as shown 
in color bars in top region and in right region. Horizontal red line marks $R_\sigma=0.69$ for the \obj. Vertical 
red line marks flux ratio 3.75 of observed broad H$\alpha$ to observed broad H$\beta$ for the \obj.
}
\label{model}
\end{figure*}

	Then, considering one of the two shifted broad components in broad H$\alpha$ is obscured with Balmer 
decrement $BD$ from 2.8 to 30 from a normal distribution with a mean of 2.8 (intrinsic flux ratio of broad 
H$\alpha$ to broad H$\beta$) and a standard deviation of 10, leading the created broad H$\beta$ to be described as 
\begin{equation}
\begin{split}
k~\in&~random(0,~1) \ \ \ \ \ \ \ \  sca~=~6564.61/4862.68 \\
f_\lambda(H\beta)~=&~G(\lambda,~[\frac{\lambda_{0b\alpha}}{sca},
	~\frac{\sigma_{0b\alpha}}{sca},~\frac{f_{0b\alpha}}{2.8}])\\
	&~+~G(\lambda,~[\frac{\lambda_{0r\alpha}}{sca},~\frac{\sigma_{0r\alpha}}{sca},
	~\frac{f_{0r\alpha}}{BD}])\ \ \ \ \ (k~<~0.5) \\
f_\lambda(H\beta)~=&~G(\lambda,~[\frac{\lambda_{0b\alpha}}{sca},
	~\frac{\sigma_{0b\alpha}}{sca},~\frac{f_{0b\alpha}}{BD}])\\
	&~+~G(\lambda,~[\frac{\lambda_{0r\alpha}}{sca},~\frac{\sigma_{0r\alpha}}{sca},
	~\frac{f_{0r\alpha}}{2.8}])\ \ \ \ \ (k~>~0.5)
\end{split}
\end{equation}
where $k$ is a random value selected from 0 to 1 to randomly determine which shifted component (blue-shifted 
or red-shifted) is obscured, 4862.68\AA~ and 6564.61\AA~ are the theoretical central wavelengths of broad 
H$\beta$ and broad H$\alpha$ in rest frame, and $\lambda$ represents the rest wavelength from 4400\AA~ to 
5400\AA~ for the simulated broad H$\beta$. Here, application of a normal distribution of $BD$ is mainly due 
to consideration that numbers of Type-1.9 AGN with seriously obscurations on central BLRs are 
apparently smaller than the number of normal Type-1 AGN with no obscurations on central BLRs as simply shown 
in \citet{ok15}\footnote{Actually, part of the newly found Type-1 AGN in \citet{ok15} are Type-1.9 AGN with 
apparent broad H$\alpha$ but no apparent broad H$\beta$.}. In order to show effects of distributions of 
$BD$ on final results, uniform distribution of $BD$ from 2.8 to 30 is also considered in the section.

\begin{figure}
\centering\includegraphics[width = 8cm,height=10cm]{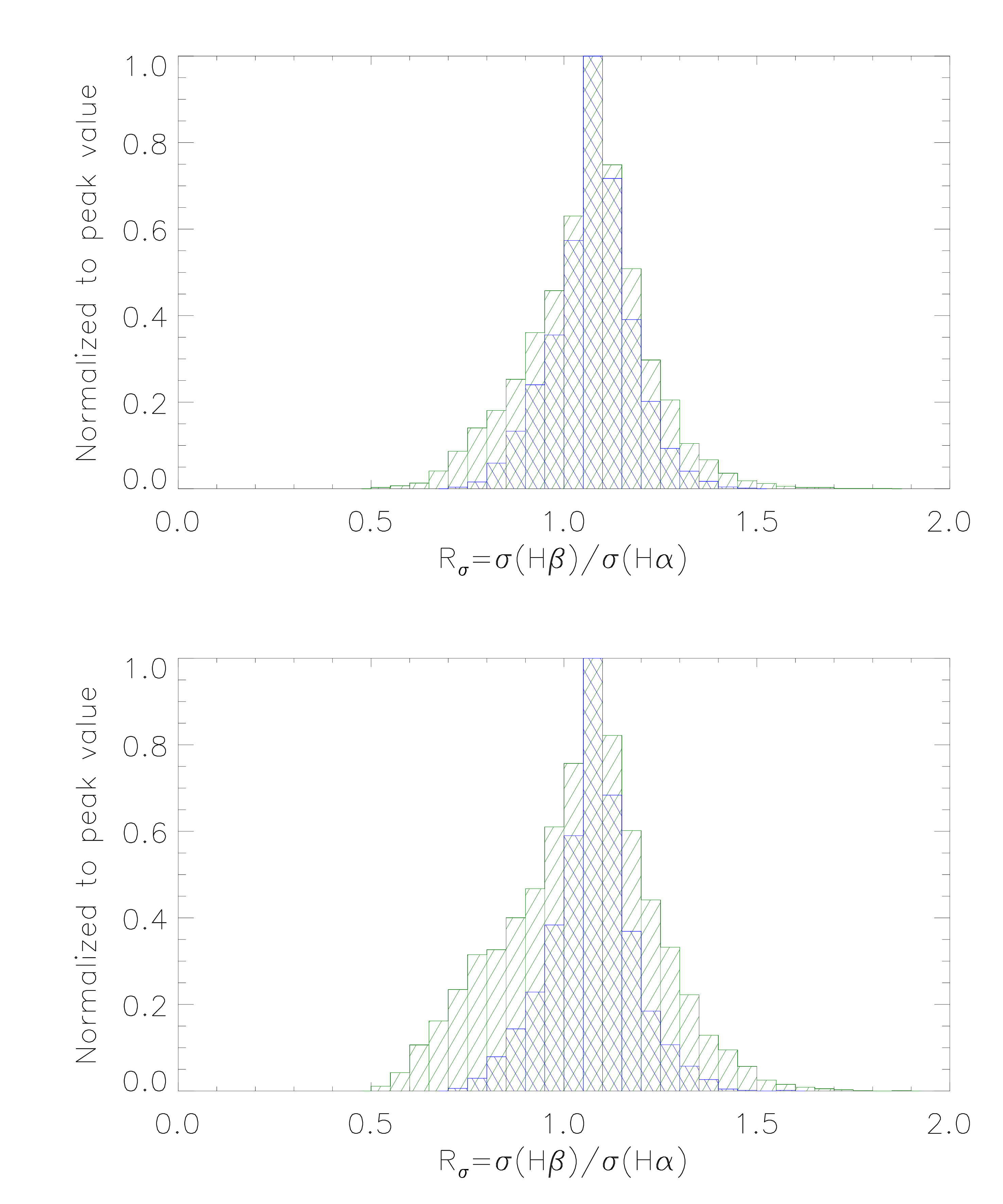}
\caption{Distributions of $R_\sigma$ through different ranges of BDs. In top panel, histogram filled  
with dark green lines and with blue lines show the results with normal distributions of $2.8~<~BD~<~30$ and of 
$2.8~<~BD~<~10$, respectively. In bottom panel, histogram filled with dark green lines and with blue lines show  
the results with uniform distributions of $2.8~<~BD~<~30$ and of $2.8~<~BD~<~10$, respectively.
}
\label{mo2}
\end{figure}

	For each simulated broad H$\alpha$ and broad H$\beta$, second moment $\sigma_0$ can be calculated 
by definition
\begin{equation}
\sigma_0^2~=~\frac{\int\lambda^2f_\lambda\dif\lambda}{\int~f_\lambda\dif\lambda}~-~
	(\frac{\int\lambda f_\lambda\dif\lambda}{\int~f_\lambda\dif\lambda})^2 \\
\end{equation}
with $f_\lambda$ as line profile of broad H$\alpha$ (broad H$\beta$). Then, after 10000 simulations under 
the assumed oversimplified BBH systems, there is an interesting diagram of second moment (in unit of ${\rm km/s}$) 
ratio $R_\sigma=\sigma(H\beta)/\sigma(H\alpha)$ versus flux ratio of simulated broad H$\alpha$ to simulated 
broad H$\beta$, shown in Fig.~\ref{model}, indicating that effects of obscurations on central two independent 
BLRs related to a BBH system can lead to quite different broad Balmer emission lines with $R_\sigma$ quite 
different from 1.1. Meanwhile, in order to check effects of different BDs on final results, 
Fig.~\ref{mo2} shows distributions of $R_\sigma$ based on different ranges of BDs. It is clear that larger 
ranges of BDs can lead $R_\sigma$ more apparently different from 1.1, as what we can expect. Moreover, the 
\obj~ with $R_\sigma=0.69$ and with flux ratio 3.75 of broad H$\alpha$ to broad H$\beta$ is not an isolated 
object, and there should be a sample of objects with quite different broad Balmer emission lines, accepted 
BBH systems common in central regions of galaxies. Unfortunately, there is no clear information on distribution 
of Balmer decrements of broad Balmer lines, therefore, there are no further discussions in current stage on 
distributions of objects with different $R_\sigma$.

	Furthermore, as shown in Fig.~\ref{model}, even for Type-1 AGN with flux ratio around 4 of broad H$\alpha$ 
to broad H$\beta$, there could be quite different line widths between broad Balmer emission lines. Meanwhile, 
as shown in Fig.~\ref{model} with normal distributions of BD and uniform distributions of BD, about 1.04\% and 3.89\% of 
the simulated results with $R_\sigma$ smaller than 0.69. Considering total 18600 normal SDSS QSOs \citep{rg02, ro12, 
pr15, lh20} with redshift smaller than 0.3 in DR16, more than 260 SDSS QSOs will be expected to have quite different 
line profiles of broad Balmer emission lines, similar as the case in \obj. Therefore, to report and study a sample 
of Type-1 AGN with quite different broad Balmer emission lines is one of our main tasks in the near future, which will 
provide further clues to determine whether different properties of broad Balmer emission lines can be accepted as 
efficient signs for BBH systems.

	Before end of the manuscript, there are two noteworthy points. On the one hand, \obj~ has been included 
in the sample of broad line AGN in \citet{ok15} with broad H$\alpha$ having reported full width at half maximum 
(FWHM) about 1400${\rm km/s}$ which is about 1.86 times smaller than the measured FWHM=2600${\rm km/s}$ of broad 
H$\alpha$ in the manuscript. The different results strongly indicate that one broad Gaussian function plus three 
narrow Gaussian functions are not appropriate to describe emission lines of H$\alpha$ and [N~{\sc ii}] doublet in 
\obj~ as what have been done in \citet{ok15}. In other words, in order to find more reliable measured line 
parameters of broad emission lines, complicated but more preferred model functions should be carefully applied 
at least in some broad line AGN, similar as what have been done in \obj~ in the manuscript. 
On the other hand, based on the results shown in the manuscript, a being prepared sample of SDSS QSOs with quite 
different line profiles of broad Balmer emission lines and corresponding clues to support optical QPOs in the 
collected QSOs should be reported in the near future.

\section{Summaries and Conclusions}

	The final summary and main conclusions are as follows. 
\begin{itemize}
\item Motivated by different broad Balmer emission lines as signs of BBH systems, \obj~ is reported 
	and discussed in the manuscript, due to its quite different line widths of broad Balmer 
	emission lines.
\item After subtractions of SSP method determined host galaxy contributions, multiple Gaussian functions 
	are applied to measure line parameters of broad Balmer emission lines in \obj, leading the broad 
	H$\beta$ having its line width (second moment) 760${\rm km/s}$ to be only about 0.69 times of 
	line width (second moment) 110${\rm km/s}$ of the broad H$\alpha$. However, common line width 
	ratio of broad H$\beta$ to broad H$\alpha$ is about 1.1 with scatter of 0.1dex in quasars.
\item Different model functions have been applied to measure the broad H$\alpha$, providing strong evidence 
	to support that the quite broader component in broad H$\alpha$ can be confirmed and preferred 
	with confidence level higher than $5\sigma$.
\item In order to naturally explain the quite different broad Balmer emission lines in \obj, a BBH system 
	can be well applied, after considering quite different obscuration effects on central two 
	independent BLRs, leading to lack of corresponding broader component in H$\beta$.
\item Strong clues are reported on BBH system expected optical QPOs with periodicity about 1000days 
	through long-term variabilities and corresponding phase-folded light curves from ZTF which 
	can be well described by a sinusoidal function plus a linear trend. And the GLS power properties 
	provide confidence level higher than $3\sigma$ to support the optical QPOs with periodicity around 
	1000days.
\item Through CAR process simulated light curves to trace intrinsic AGN activities, confidence level higher 
	than $2\sigma$ can be confirmed to support that the optical QPOs from ZTF light curves in \obj~ are 
	not from intrinsic AGN activities, although the collected ZTF light curves have short time durations 
	only 1.47 times of the expected periodicity of the QPOs.
\item Accepted BBH systems leading to quite different broad Balmer emission lines with considering different 
	obscurations on central two BLRs, simulated results can provide clues to report and discuss more broad 
	line quasars with quite different broad Balmer emission lines but with normal flux ratios around 4 
	of broad H$\alpha$ to broad H$\beta$ in the near future, to clearly test whether different properties 
	of broad Balmer emission lines can be accepted as efficient signs for BBH systems in normal broad 
	line AGN.
\end{itemize}

\section*{Acknowledgements}
%Zhang gratefully acknowledges the anonymous referee for reading our manuscript carefully 
%and patiently.
Zhang gratefully acknowledge the anonymous referee for giving us constructive comments and suggestions 
to greatly improve our paper. Zhang gratefully acknowledges the kind grant support from NSFC-12173020. This paper 
has made use of the data from the SDSS project, \url{http://www.sdss3.org/}, managed by the Astrophysical Research 
Consortium for the Participating Institutions of the SDSS-III Collaboration. This paper has made use of the data 
from the CSS \url{http://nesssi.cacr.caltech.edu/DataRelease/} and the data from the ZTF \url{https://www.ztf.caltech.edu}, 
and the data from the ASAS-SN \url{https://asas-sn.osu.edu/}. The paper has made use of the public JAVELIN code 
\url{http://www.astronomy.ohio-state.edu/~yingzu/codes.html#javelin}, and the MPFIT package 
\url{https://pages.physics.wisc.edu/~craigm/idl/cmpfit.html}, and the emcee package 
\url{https://emcee.readthedocs.io/en/stable/}. This research has made use of the NASA/IPAC Extragalactic Database 
(NED, \url{http://ned.ipac.caltech.edu}) which is operated by the California Institute of Technology, under contract 
with the National Aeronautics and Space Administration.

\section*{Data Availability}
The data underlying this article will be shared on reasonable request to the corresponding author
(\href{mailto:aexueguang@qq.com}{aexueguang@qq.com}).

\label{lastpage}
\end{document}